\documentclass[letterpaper,twocolumn,10pt]{article}

\usepackage{usenix}
\usepackage{inconsolata}
\usepackage[T1]{fontenc}
\usepackage{pgfplots}
\usepackage{graphicx}
\usepackage{todonotes}
\usepackage[dvipsnames]{xcolor}
\usepackage{makecell}
\usepackage{tikz}
\usepackage[ruled,linesnumbered]{algorithm2e}
\usepackage{graphicx,listings,tikz}
\usepackage{pgfplots}
\usepackage{wrapfig}
\usepackage{makecell}
\usepackage{caption}
\usepackage{hyperref}
\usepackage{subcaption}
\usepackage{calc}
\usepackage{xfp}
\usepackage{float}
\usepackage{amsmath,amsthm,amssymb}
\usepackage{tikz-cd, amssymb}
\usepackage{mathrsfs}
\usetikzlibrary{shapes.geometric}

\tikzcdset{arrow style=tikz, diagrams={>={Triangle[open, length=2mm]}}}

\theoremstyle{definition}
\newtheorem{example}{Example}
\definecolor{goalcolor}{rgb}{0.64, 0.0, 0.0}

\usetikzlibrary{shapes.geometric,arrows,fit,calc,positioning,automata,backgrounds,fit,patterns}
\usetikzlibrary{shapes,shapes.geometric,arrows,fit,calc,positioning,automata}
\tikzset{eli state/.style={draw,ellipse}}%fill={rgb:black,1;white,10}
\tikzset{rect state/.style={draw,rectangle}}%,fill={rgb:black,1;white,10}
\tikzset{diamond state/.style={draw,diamond}}%,fill={rgb:black,1;white,10}
\tikzset{blank state/.style={draw=none}}
\newcounter{barcount}
\tikzset{barchart/.cd,y distance/.initial=3em,
bar height/.initial=2em,width/.initial=6cm,
bar text/.style={font=\sffamily,text depth=0.25em},
description/.style={font=\sffamily,text depth=0.25em},
colors/.initial={"orange!80","blue!40","red","green!70!black"}}

\def\addlegendimage{\csname pgfplots@addlegendimage\endcsname}

\newcommand\score[2]{%
  \pgfmathsetmacro\pgfxa{#1 + 1}%
  \tikzstyle{scorestars}=[star, star points=5, star point ratio=2.25, draw, inner sep=0.15em, anchor=outer point 3]%
  \begin{tikzpicture}[baseline]
    \foreach \i in {1, ..., #2} {
      \pgfmathparse{\i<=#1 ? "black" : "white"}
      \edef\starcolor{\pgfmathresult}
      \draw (\i*1em, 0) node[name=star\i, scorestars, fill=\starcolor]  {};
    }
    \pgfmathparse{#1>int(#1) ? int(#1+1) : 0}
    \let\partstar=\pgfmathresult
    \ifnum\partstar>0
      \pgfmathsetmacro\starpart{#1-(int(#1)}
      \path [clip] ($(star\partstar.outer point 3)!(star\partstar.outer point 2)!(star\partstar.outer point 4)$) rectangle 
      ($(star\partstar.outer point 2 |- star\partstar.outer point 1)!\starpart!(star\partstar.outer point 1 -| star\partstar.outer point 5)$);
      \fill (\partstar*1em, 0) node[scorestars, fill=black]  {};
    \fi
  \end{tikzpicture}%
}

\pgfplotsset{ 	ybar/.append style={ area legend} }
\def\addlegendimage{\csname pgfplots@addlegendimage\endcsname}

\def\tsc#1{\csdef{#1}{\textsc{\lowercase{#1}}\xspace}}
\tsc{WGM}
\tsc{QE}
\tsc{EP}
\tsc{PMS}
\tsc{BEC}
\tsc{DE}
\begin{document}
\title{\Large \bf Automated Generation of Cybersecurity Exercise Scenarios}

%\author{
%{\rm Anonymous Authors}}

%for single author (just remove % characters)
\author{
{\rm Charilaos Skandylas}\\
Linköping University
\and
{\rm Mikael Asplund}\\
Linköping University
}

\lstdefinelanguage{Topology}{
    morekeywords=[1]{System, Topology, Subsystem, Constraint},
                  morekeywords=[2]{},
    sensitive=false, % keywords are not case-sensitive
    morecomment=[l]{//}, % l is for line comment
    morecomment=[s]{/*}{*/}, % s is for start and end delimiter
    morestring=[b]" % defines that strings are enclosed in double quotes
} % 

\lstset {
    language=Topology,
    %frame=tb,
    %tabsize=1,
    %showstringspaces=false,
    %numbers=left,
    %upquote=true,
    commentstyle=\ttfamily\color{ForestGreen},
    keywordstyle=[1]\ttfamily\color{blue},
    keywordstyle=[2]\ttfamily\color{red},
    basicstyle=\ttfamily\footnotesize,
    %emphstyle={\color{blue}},
    %keywordstyle=\color{blue},
    %columns=relative,
    %breaklines=true,
    %basewidth = {.45em},
    %xleftmargin=0.7cm,
}

\lstdefinelanguage{ArchStyle}{
    morekeywords=[1]{ArchStyle, Role, Feature, Constraint, Interface, Map, Vulnerability, Entrypoint},
                  morekeywords=[2]{requires,},
    sensitive=false, % keywords are not case-sensitive
    morecomment=[l]{//}, % l is for line comment
    morecomment=[s]{/*}{*/}, % s is for start and end delimiter
    morestring=[b]" % defines that strings are enclosed in double quotes
} % 

\lstset {
    language=ArchStyle,
    %frame=tb,
    %tabsize=1,
    %showstringspaces=false,
    numbers=left,
    %upquote=true,
    commentstyle=\ttfamily\color{ForestGreen},
    keywordstyle=[1]\ttfamily\color{blue},
    keywordstyle=[2]\ttfamily\color{red},
    basicstyle=\ttfamily\footnotesize,
    %emphstyle={\color{blue}},
    %keywordstyle=\color{blue},
    %columns=relative,
    %breaklines=true,
    %basewidth = {.45em},
    %xleftmargin=0.7cm,
}

\lstdefinelanguage{Alloy}{
    morekeywords=[1]{var, enum, abstract, sig, init, bool, true, false, one, lone, all, some, set, reward, implies,  run, formula, pred, iff, or, not, in, fact},
     morekeywords=[2]{</, />},
    sensitive=false, % keywords are not case-sensitive
    morecomment=[l]{//}, % l is for line comment
    morecomment=[s]{/*}{*/}, % s is for start and end delimiter
    morestring=[b]" % defines that strings are enclosed in double quotes
} % 

\lstset {
    language=Alloy,
    %frame=tb,
    %tabsize=1,
    %showstringspaces=false,
    numbers=none,
    %numbersep=0em,
    %upquote=true,
    commentstyle=\ttfamily\footnotesize\color{ForestGreen},
    keywordstyle=[1]\ttfamily\footnotesize\color{blue},
    keywordstyle=[2]\ttfamily\footnotesize\color{red},
    basicstyle=\ttfamily\footnotesize,
    %emphstyle={\color{blue}},
    %keywordstyle=\color{blue},
    %columns=relative,
    %breaklines=true,
    %basewidth = {.45em},
    %xleftmargin=0.7cm,
}

\maketitle
\begin{abstract}
There is a growing need for cybersecurity professionals with practical knowledge
and experience to meet societal needs and comply with new standards and regulations.
At the same time, the advances in software technology and artificial intelligence
point towards a future where software agents will play an important role in protecting
the computer systems that are critical for society to function.
The training and development of both humans and software agents requires the design and execution of cybersecurity exercises that differ in properties such as size, scope, objectives, difficultly, etc.
Cybersecurity scenarios are critical for the operation of cybersecurity exercises as they describe the scope, context, operational environment and storyline of each exercise. 
 In this work, we present an approach to automatically generate cybersecurity scenarios that model enterprise IT systems. Our approach is able to generate a large number of scenarios that differ in multiple criteria including size, scope, difficulty, complexity and diversity. We further release as open source: a simulation and a virtualization environment that can run cybersecurity exercises based on the generated 
scenarios and a dataset containing 100000 sample scenarios.
\end{abstract}
\section{Introduction}
The security of computer systems has become a primary concern for organizations, businesses, individuals and even national security.
For instance, cybersecurity has become a top priority in governmental services, critical infrastructures, banking and payments, healthcare.
Maintaining a {good level} of security is paramount for the functioning of society with respect to the safety and privacy of individuals and the
non-interrupted {functioning of governmental institutions}. However, cybersecurity incidents over the last decade including NotPetya,
Solarwinds and multiple ransomware attacks on critical infrastructures and banking, e.g., WannaCry, Colonial Pipeline, and JBS foods indicate a significant gap in
operational security and the ability to react in time to and mitigate cybersecurity threats. It is evident therefore that a significant amount of training is required
to improve the cyberdefense capabilities of organizations including the cybersecurity operators and {ordinary} personnel~\cite{BRILINGAITE2020101607,katsantonis2023cyber}. At the same time, recent
research efforts have been directed towards (semi-)autonomous agents that can aid in cyberdefense operations. These agents also require a significant amount of training
to be effective~\cite{10.1007/978-3-031-54129-2_43}. 
The principal means of training for operational security both in terms or readiness and mitigation  has been the design and running of cybersecurity exercises. 

In this paper, we propose a method to generate cybersecurity scenarios. In this context, we consider a cybersecurity scenario to include three elements: (i) a scenario execution
environment, i.e., the platform and computational elements that the scenario will run on, (ii) a set
of roles and corresponding actors that play those out, and (iii) a storyline that defines how the
scenario is allowed to play out, including the goals, available actions and objectives of each
participating actor. A scenario may additionally provide the foundations for other systems required
for a cybersecurity exercise, for instance: provide a scoring mechanism to calculate competition scores
or provide rewards to facilitate the construction of training sessions or running competitions.

There exist a number of scenario modeling approaches in the literature.
The most popular approach is the use of scenario definition languages~\cite{CST-SDL,10.1007/978-3-030-42051-2_8,VSDL,Weiss,KYPO,CyRIS} where all scenario elements are explicitly specified in a machine-readable format. Alternative scenario modeling approaches include serious games~\cite{SERIOUS}.
The above  scenario modeling approaches involve a high-degree of human interaction and a
low degree of design automation, contrary to the almost fully automated scenario deployment. The designer needs to define and select each an every individual element to be included in the scenario, a process that is tedious and error-prone requiring significant amount of effort, which in turn affects both the realistic number of varied scenarios that can be designed and their scalability. Automated scenario generation can reduce human errors and the required effort thus improving  scalability, while, scenario security modeling at the architectural level, where the details of the scenario element generation can be automatically filled with different viable alternatives can
increase the variety of available scenario elements. 
Moreover, most scenario generation approaches focus on either simulated or emulated
environments. In practice, cybersecurity exercise deployment environments take multiple forms, from tabletop exercises using pen and paper to fully integrated emulation or virtualization environments to real hardware-backed networks, therefore, a scenario generation approach that can support multiple types of execution environments seamlessly allows scenario re-use across different exercise running organizations irrespectively of their available resources.
An often requested setup for the development of ML/RL agents that require large amounts of executions is that training is performed on the less costly simulated environment and the evaluations of effectiveness and performance are performed in the more realistic virtualized environment.

A small number of works have attempted to achieve the same goal as this paper, 
Cyexec$^{*}$~\cite{nakata2021cyexec} employs graph randomization techniques to rearrange a number of predefined scenario segments to provide a large number of different scenarios.
It generates a container based environment on which the exercises can run.
  Even though Cyexec$^{*}$ is able to generate a large number of scenarios, since the scenario fragments are pre-determined the opportunity to learn from the different scenarios is limited.
In this work, we aim to go beyond randomization of a set of fixed 
scenario fragments. Instead of shuffling scenario fragments, 
we generate a large amount of viable alternative subsystems that can are freely
interchangeable and can be composed using different network topologies to generate
scenarios that differ both in their elements and their structure. We furthermore,
randomize the in-subsystem objectives and as a result can provide  a larger degree
of variation compared to Cyexec$^{*}$. Large language models~\cite{10695083} and machine learning~\cite{zacharis2023ai,zacharis2023aicef}
have also used to automatically generate scenarios, however, these two approaches do not automatically generate an execution environment or a machine readable specification of 
a scenario. They rather run similar to tabletop exercises and would require additional effort to clarify the details required to generate a machine readable specification and 
further effort to automate their deployment.

In~\cite{RUSSO2020101837}, Russo et.al  define a list of six desired properties for cybersecurity scenario generation
approaches. Given the previously identified needs for automation and variety, we extend the list as follows:
\textbf{1)} extensibility (E), i.e., the ability to support the seamless integration
of new elements (e.g., vulnerabilities, software, and hardware).
\textbf{2)} modularity (M), i.e., the ability to reuse and compose elements defined within the framework
  without customizing them,
\textbf{3)} solvability (So), which we define as the combination of (i) verifiability, i.e., the ability
  to validate the scenario environments and (ii) testability, the ability to test the scenario execution,
\textbf{4)} scalability (Sc), the ability to support scenarios or different sizes and scale with them
\textbf{5)} difficulty (D), i.e., the ability to generate scenarios that vary in difficulty,
  in other words that take varying amounts of effort to complete,
\textbf{6)} {variety (V)}, the ability to generate different types of networks,
  including different network elements, different topologies and domains,
\textbf{7)} compatibility, i.e., the usage of standard/well-established and widely supported infrastructure development technologies.

%\todo[inline]{Missing: Systems driven to be able to model different networks and components, ability to define components with different viable interfaces }
In this paper, we propose an automated approach for cybersecurity scenario generation.
Our approach has two phases, execution environment generation and storyline generation.
We employ model finding to generate a large number of valid and {interesting}
execution environments. Each execution environment represents a network of subsystems, each comprising a number of exploitable components, typically network hosts, with vulnerable interfaces, usually
services or processes). Subsystems are connected to each other through connectors, typically routers,
which also expose potentially exploitable interfaces. Connectors can also act as bridges and firewalls.
Our storyline generation approach is inspired by quest generation approaches in video games~\cite{yu2021towards,ashmore2007quest}.
We employ a \emph{lock and key} generation approach to introduce \emph{challenges} that need to be
solved before the red-team agent can succeed in the scenario \emph{objectives}.
The scenario \emph{objectives} are generated by considering the network contents and the actions available to the red-team agents. 
We have implemented and release as open source: a scenario generator that follows our proposed approach to generate scenarios and two scenario backend platforms 
on which cybersecurity exercises can run. The first backend is a simulation-based environment, similar to environments like NaSIM and  CyBORG used to train autonomous red or blue agents, and the second target is a libvirt-based emulation environment able to generate a realistic VM network.

In summary, this paper contributes:
\begin{itemize}
\item An automated approach to generate cybersecurity scenarios with the identified desired properties,
\item A scenario generation toolset, including a scenario generator, an environment and storyline simulation backend and libvirt-based virtualization backend.
\item An evaluation of our approach in terms of the identified properties and it's performance.
\item A dataset containing 100000 networks.
\end{itemize}

The rest of this paper is structured as follows:
Section~\ref{sec:preliminaries} provides a high-level introduction to architectural modeling for security, to model finding and alloy.
Section~\ref{sec:overview} gives a high-level overview of our approach.
Section~\ref{sec:execenv} discusses how we generate the execution environment of a scenario.
Section~\ref{sec:storyline} discusses how we generate the scenario storyline.
Section~\ref{sec:scenariorealization} discusses how our scenarios can be used in a simulated and a virtualized environment to run cybersecurity exercises.
Section~\ref{sec:evaluation} details the evaluation of our approach.
In Section~\ref{sec:related.work}, we discuss and situate ourselves with regards to related work.
Section~\ref{sec:conclusion} concludes this paper.

\section{Preliminaries}
\label{sec:preliminaries}
\subsection{Architectural Modeling for Security}
We define a software architecture as a composition of (i) components, (ii) interactions and (iii) architectural properties.
{Components specify the behavior of individual parts of the
system. Each component comprises a set of interfaces that can be invoked 
to provide the system functionality associated with them. Each interface
is associated with a set of capabilities. A capability is the ability to perform a certain set of actions on a component or its environment.
{Capabilities are used to allow flexibility in describing  the effects of actions in systems that belong to different domains.}
{Examples of capabilities include: calling a remote API, initiating a remote connection, adding/removing an interface or component, creating/opening/closing a file, running a child process, spawning a thread etc.}
An interaction is an invocation of one or more target interfaces from a source interface to realize the corresponding functionality.
An architectural property represents an important quality about a component or an interface.
{
A component property refers to relevant information related to the component as a
whole that is not specific to an interface. For instance, if we assume that the component
represents a network host, the host’s operating system or IP address would constitute such
properties. Interface properties constitute information that is specific to interfaces.
For instance, if an interface refers to a network service, a relevant property would be the port on
which the service is available, if the interface refers to a process, a relevant
property would be the process identifier.}

A security-informed architecture is a software architecture whose properties concern security-related information including the vulnerabilities associated with each component, the interfaces that can be invoked to exploit those vulnerabilities and how vulnerabilities can be combined to form complex attacks that allow the attacker to compromise the system. %We consider a vulnerability~\cite{236681} to be: "An error, flaw, or mistake in computer software that permits or causes unintended behavior to occur".

\subsection{Model Finding}
Finite model finding in first-order logic refers to finding all models that satisfy 
a set of first-order logic formulas up to a size bound.
In the MACE-style approach~\cite{janota2018towards}, a domain size for the finite model is selected, followed by a grounding of the first-order problem with this domain and translating the resulting formulas into a SAT problem. If the SAT problem is satisfiable, each solution is a finite model of the selected size. In this work we use a related tool, Alloy for model
finding.

\paragraph{Alloy}
Alloy's logic is an extension of first-order logic and is based on relations.
Functions are modeled as binary relations and sets are modeled as unary relations. A specification reduces to a set of first-order constraints
over a set of relational variables. Each variable is a tuple represented by a set of tuples, the values of which are drawn from a universe of atoms, i.e., uninterpreted elements.
An alloy model is a collection of signatures, relations and facts.
A signature represents a typed set with an arity and is associated with a set or relations with other signatures. Predicates are boolean functions and facts correspond to axioms that constraint the possible variable instantiations.

An example alloy specification is shown in Figure~\ref{fig:sigdef}
It defines four signatures. The Component signature has a multiplicity of zero and
defines a relation inSystem that maps any instantiation of a Component to an instantiation
 of a single system.  Two component subtypes are defined as Backend and Frontend with a multiplicity of at most one and inheriting system from Component. The System signature has a multiplicity of exactly one and each of its instantiations (only one can be possible) maps it to a least one or more components. The predicate requires states that 
 for two component instantiations, if one has more than zero instances in the model so does the other and vice versa. Additionally, if one instantiation is a member of System.components then the other one is as well. Finally, is specifies that the elements of inSystem are the same for both c0 and c1.  
 The systemcomposition fact asserts: (i) all instances of type Backend (zero or one) are a member of the System.components relation and (ii) that Backend requires Frontend.
  Only one model satisfies the specification, it includes three atoms, System, Backend and Frontend,
 components(System) = \{ Backend, Frontend \} and system(Backend) = \{ System \} and system(Frontend) = \{ System \}.
\begin{figure}
\begin{lstlisting}[language = Alloy]
abstract sig Component{inSystem: one System}
lone sig Backend extends Component {}
lone sig Frontend extends Component {}
one sig System {components: some Component}
pred requires[c0: Component, c1:component]{
#c0 > 0 iff #c1 > 0
c0 in System.components <=> c1 in System.components
c0.inSystem = c1.inSystem }
fact systemcomposition{
Backend in System.components
requires[Backend,Frontend] }
\end{lstlisting}
\caption{Example alloy specification}
\label{fig:sigdef}
\end{figure}

\section{Scenario Generation Overview}
\label{sec:overview}
%We model a network as a system of systems, i.e., as a set of \emph{subsystems} that are connected to
%each other via a set of \emph{connectors}. Subsystems implement the network's intended functionality while
%connectors facilitate communication between the different subsystems but also act in other
%extra-functional capacities, e.g., as firewalls, load balancers, etc.
%Each subsystem is further modeled as a set of communicating
%\emph{components}. Both components and connectors can have vulnerabilities that can be exploited by
%an adversary to gain capabilities over the system. Examples of capabilities include bypassing a connector's firewall
%rules, taking over a component, redirecting connections etc. 

A high-level overview of our approach to generate cybersecrurity scenarios can be seen in Fig.~\ref{Fig:HighLevelApproach}.
We generate scenarios in two phases: first we generate the execution environment that scenarios will run on  and second we generate the scenario's storyline.
The execution environment generator receives as input: (i) a target network topology specification (ii) a number of target subsystems (iii) a set of constraints among the subsystems. Further, the scenario generator depends on
a repertoire of architectural style specifications to 
generate the subsystem elements. 
The storyline generator in addition to the output of the execution environment generator additionally
receives a set of configuration parameters that enable the tuning of the scenario properties, e.g.,
difficulty. In the general case, for a single input set, the execution environment generator will produce
a large number of viable networks with the target network architecture that contain all the required
subsystems that conform to the specified constraints. For each of the generated networks,
with the same set of configuration parameters, a large amount of scenarios can be generated.

\begin{figure}
  \centering
  \includegraphics[scale=0.45]{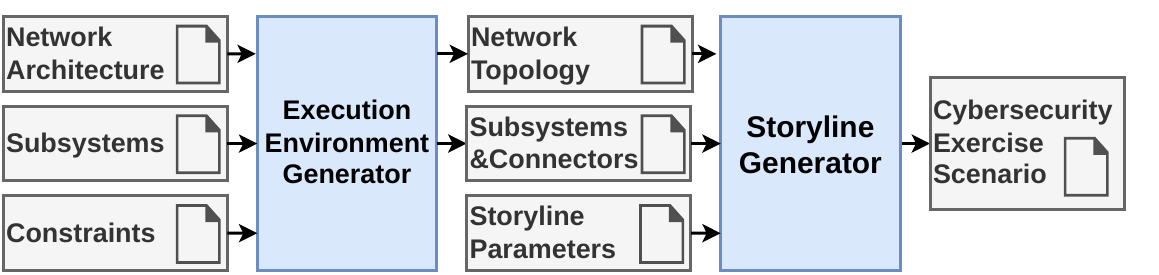}
  \caption[justification=centering]{Scenario Generation Overview}
  \label{Fig:HighLevelApproach}
\end{figure}

\section{Execution Environment Generation}
\label{sec:execenv}
In this section we discuss the cybersecurity scenario environment generation process which is depicted
in Fig.~\ref{Fig:EnvironmentGeneration}
The process comprises two parts: (i) the generation of the network topology that comprises connectors and subsystems and (ii) the detailed definition of the connectors and subsystems.

The execution environment generator receives as input a network specification which comprises the following elements: (i) the target network architecture, (ii) the architectural style and a number of each subsystem to be included in the network and (iii) a set of constraints between the different subsystems. It produces two outputs that make up 
the network environment: (i)
a network topology that expresses
how the connectors and subsystems
are connected to each other,
and (ii) a detailed representation
of each connector and subsystem
in the environment
which can then be simulated or used to create a number of virtual machines.

\begin{figure}
  \centering
  \includegraphics[scale=0.7]{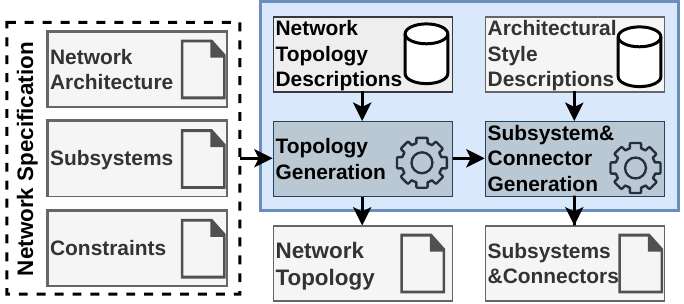}
  \caption[justification=centering]{Execution Environment Generation Process}
  \label{Fig:EnvironmentGeneration}
\end{figure}

\subsection{Topology Generation}
Figure~\ref{fig:topologyspec} shows an example network specification that defines a \emph{collapsed core network}~\cite{paguigan2022assessment} (line 1), with three subsystems: \emph{ClientServer}, \emph{Microservices} and \emph{OAuth}, (lines 2-4). It additionally defines
two relational constraints, one between Oauth and Microservices and one between Oauth and ClientServer.

\begin{figure}
  \centering
\begin{lstlisting}[language = Topology]
Topology CollapsedCore
Subsystem ClientServer:1
Subsystem Microservices:1
Subsystem OAuth:1
Constraint requires OAuth Microservices
Constraint requires OAuth ClientServer
Constraint allowRouting Microservices ClientServer
\end{lstlisting}
\caption{An example topology definition}
\label{fig:topologyspec}
\end{figure}

We support multiple constraint relations among subsystems including: co-existence and comes-before and routing(allowRouting) and/or firewall rules.
Additional constraint relations can be encoded in first order logic and provided to the environment generator as part of the network specification. 
Routing rules are specified through the traffic predicate.

We have implemented an alloy general description of valid network topologies featuring subsystems and connectors to which we append a translation
to alloy, of the network specification to produce 
a customized alloy model that generates all valid network architectures that satisfy both the general description of valid topologies and the specific network specification. The general description of valid network topologies can be found in appendix~\ref{Alloy}.
To translate the network specification to an alloy 
specification we follow the following steps:
\textbf{1)} for every subsystem, we generate a new signature for the subsystem's architectural style and add a fact that specifies the
number of instances to generate.
\textbf{2)} We add a predicate for every subsystem constraint.

Figure~\ref{fig:alloytopologyspec} shows the alloy specification generated from the network specification in Figure~\ref{fig:topologyspec}. Lines 1-3 and 9-11 are the results of step \textbf{1)} and lines 5 and 6 are the results of step \textbf{2)}.

 \begin{figure}
  \centering
\begin{lstlisting}[language = Alloy]
one sig ClientServer extends ArchStyle { }
one sig Microservices extends ArchStyle { }
one sig Oauth extends ArchStyle { }
fact constraint_mapping{
requires[ClientServer, Oauth]
requires[Micorservices,Oauth] }
run {
#ClientServer.implementedIn = 1
#Microservices.implementedIn = 1
#Oauth.implementedIn = 1 }
\end{lstlisting}
\caption{Alloy translation of Figure~\ref{fig:topologyspec}}
\label{fig:alloytopologyspec}
\end{figure}

The result of invoking alloy is the set of all valid 
network topologies that satisfy the specification.
A network architecture defines the valid ways in which the topology elements can connect to each other.
We currently support two target network topologies: (i) the flat network topology and (ii)
the collapsed-core network topology. The flat network topology only has a single connector to which all subsystems
are connected. A collapsed core topology~\cite{paguigan2022assessment}, is a 2-layer network topology where the core and distribution functions of the network elements, in our model represented by connectors, are placed in the same layer. Fig.~\ref{fig:network.topologies} shows two example network architectures that satisfy the specification in Figure~\ref{fig:alloytopologyspec}.

We model the topology of a network architecture as an undirected graph $ G_{T}$.
Each of the graph's nodes either represents a subsystem, i.e., $s \in S$ or a connector, i.e., $c \in C$. With $S$ representing the set of subsystems in the network architecture and $C$ representing the set of connectors.
$L$ is the set of edges, each of which represent a connector link.
Connectors can have edges to both subsystems and connectors, however, subsystems can only have edges to connectors. 
%Two alternative valid topologies generated for the
%specification in Figure~\ref{fig:topologyspec} are %shown in Figure~\ref{fig:network.topologies}.

\begin{figure}
  \centering
  \subfloat[]{\includegraphics[scale=0.4]{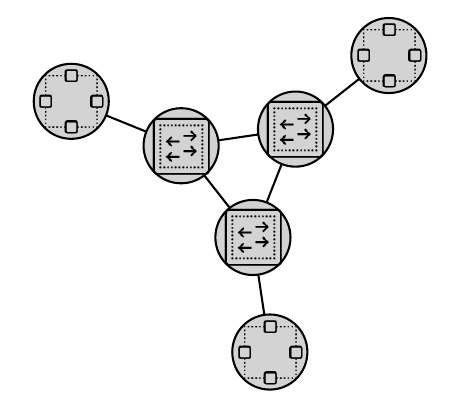}\label{fig:CC}}
  \hspace{2em}
  \subfloat[]{\includegraphics[scale=0.4]{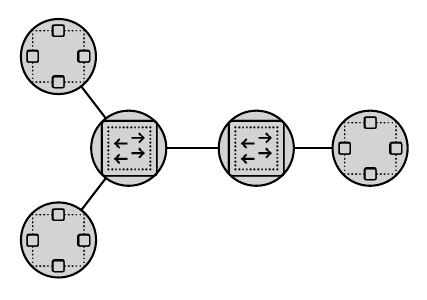}\label{fig:FN}}
  \caption{Two alternative topologies}
  \label{fig:network.topologies}
\end{figure}

\subsection{Subsystem Generation}
To generate the subsystem contents our execution environment generator parses the network definition,
identifies the subsystem architectural styles and 
looks up each style in its repertoire of style specifications.
If a specification is found for each style, it follows 
a similar process to the one taken to generate the network topology. The architectural style specification 
is parsed and translated to an equivalent alloy specification. As was the case for the topology generation, we have implemented a general description of valid architectural styles to which we 
append the architectural style specification to generate an alloy specification that can be fed to alloy to generate all valid instances of subsystems 
that follow the specified architectural style.
The general description of architectural styles can 
be found in appendix~\ref{Alloy}.

Figure~\ref{fig:archstylespec} shows part of an example 
specification of the  ClientServer architectural style. 
The specification includes 7 element types.
Architectural roles which serve as archetypes to generate components from, in addition to the number of components to generate (e.g., line 2). Features
that represent the functionality
that each role provides and therefore are tied to one or more 
roles (e.g., line 6). Interfaces which represent concrete implementations of features, usually as processes or services (e.g., line 9).
Mapping rules that define the accessibility between roles within the subsystem to be generated, 
A set of relational constraints related to roles or features (e.g., lines 11 and 13).
A set of vulnerabilities,
each of which is tied to an 
interface, in addition to the capabilities required to attempt to exploit
each vulnerability and the capabilities gained if exploitation succeeds and
finally the entrypoint(s)
i.e., the roles the generated components of which are initially reachable if the subsystem is accessible.
%It defines two component roles, \emph{WebServer},  and \emph{WebClient}. One \emph{WebServer} and 2 \emph{WebClient} components will be generated.
%There are two features, \emph{WebHosting}, and \emph{RunAs}
%associated with the WebServer role, and two features, \emph{WebAccess}, and \emph{RunAs}  associated with \emph{WebClient}.
%The \emph{Apache} interface provides the \emph{WebHosting}
%feature, the \emph{Firefox} interface provides \emph{WebAccess} and  \emph{Sudo} provides the \emph{RunAs} feature.
%All components with a \emph{WebClient} role map to the 
%single \emph{WebServer} role. 
%Three constraint relations are declared all specifying uniqueness.
%unique\_feature means that there can only be a single interface that
%provides the specified feature for the specified role and unique\_role
%means that only a single component in the architectural style can have that role. As is the case for the topology specification, additional constraint relations can be encoded in first order logic and provided to the environment generator as part of the network specification.  
%The specification lists two vulnerabilities: (i) \emph{DAWPwn}, targeting the \emph{Apache} interface, requiring no capabilities to exploit and upon exploitation conferring the \emph{InitialAccess} capability and (ii) \emph{SudoPwn} targeting the \emph{Sudo} interface,
%requiring the \emph{InitialAccess} capability to be exploited and providing
%the \emph{RootAccess} capability, i.e., root access as a result of being exploited. Finally, the specification includes the entrypoint component role which is WebServer.

 \begin{figure}
  \centering
\begin{lstlisting}[language = ArchStyle]
ArchStyle ClientServer
Role WebServer:1
Role WebClient:2
Feature WebHosting:WebServer
Feature WebAccess:WebClient
Feature RunAs:WebServer,WebClient
Interface Apache:WebHosting 0
Interface Firefox:WebAccess 0
Interface Sudo:RunAs InitialAccess
Map WebClient WebServer AllToOne
Constraint unique_feature WebServer WebHosting
Constraint unique_feature WebClient WebAccess
Constraint unique_role WebServer
Vulnerability DAVPwn Apache 0 InitialAccess
Vulnerability SudoPwn Sudo InitialAccess RootAccess
Entrypoint WebServer
\end{lstlisting}
\caption{An example subsystem architectural style definition}
\label{fig:archstylespec}
\end{figure}

The steps to translate the architectural style textual specification
to and alloy specification are similar to the ones described for the topology. In particular:
\textbf{1)} for each component role, we generate an architectural role signature and add a fact that specifies the number of component instances to generate. We additionally generate a fact that gathers all features that the role requires,
\textbf{2)} for every feature, we generate a feature signature and add a fact that gathers all interfaces that can be used to implement the feature. \textbf{3)} for every interface, we generate a signature.
\textbf{4)} for every mapping, we generate a signature and a corresponding multiplicity predicate.
\textbf{5)} for every constraint, we add a corresponding predicate.
It is important to note that not all elements of the specification are passed to alloy.
In particular, vulnerabilities, capabilities and their handling are considered separately once the subsystems have been generated.
The resulting alloy specification for the specification in Figure~\ref{fig:archstylespec} is shown in Appendix~\ref{Alloy}. %Figure~\ref{fig:alloyarchspec}.

Alloy's output for each subsystem translates to a security-informed architecture with no vulnerabilities or architectural properties.
To incorporate vulnerabilities we parse the architectural style specification once more, identify the component roles that have interfaces for which vulnerabilities are defined and add each vulnerability to the instantiated components identified interfaces.
To integrate capabilities, for every component we identify its interfaces and add an
architectural property $interactionReq$ that matches the interaction prerequisites. Similarly, to model the exploitation
prerequisite capabilities and the capabilities gained through successful
exploitation, we add two more architectural properties $exploitReqCaps$ and $exploitGainCaps$ respectively to the interface with the vulnerability. To incorporate the entrypoint, we simply add an $entrypoint$ architectural property to the component(s) that serve as entrypoints.
    
To add vulnerabilities to the connector elements we instantiate them as network routers with operating systems and services with known vulnerabilities. In essence, when it comes to vulnerability identification and management, we also treat them in the same way that we treat components, i.e., we identify each connector's interfaces and their vulnerabilities. 

\begin{figure}
  \includegraphics[scale=0.45]{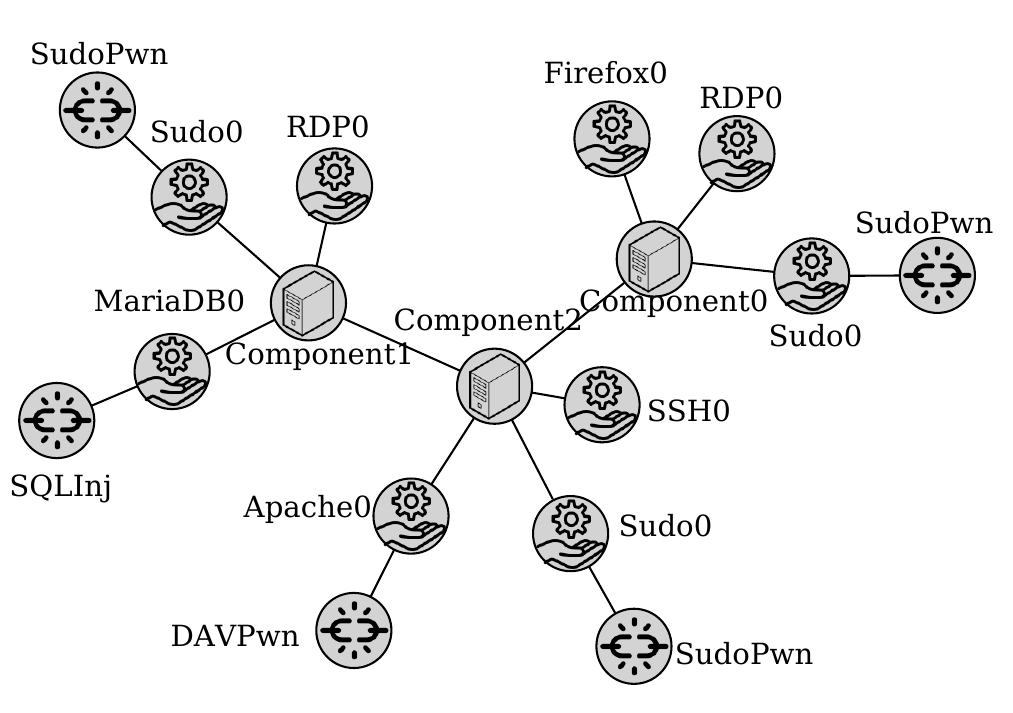}
  \caption{Example generated subsystem graph}
  \label{Fig:SubsystemGraphExample}
\end{figure}

From each security-informed architecture, we derive a subsystem graph.
A subsystem graph is an undirected graph: $G_{s}$ 
where the set of nodes is the union of $\mathcal{C}$,  $\mathcal{I}$,  $\mathcal{V}$, the sets of components, interfaces and vulnerabilities in the security-informed architecture and the set of edges comprises
the union of (i) a set of edges among component nodes $L_{\mathcal{C}}$ that interact with each other i.e., if a component $c_i$ can invoke any of another component $c_j$'s interfaces then $(c_i, c_j) \in L_{\mathcal{C}}$, (ii) a set of edges $L_{\mathcal{I}}$, from each interface to the component that contains it and (iii) a set of edges $L_{\mathcal{V}}$ from each vulnerability to the corresponding vulnerable interface.
An example subsystem graph for a ClientServer subsystem is shown in Figure~\ref{Fig:SubsystemGraphExample}.
\section{Storyline Generation}
\label{sec:storyline}
The second half of our scenario generator performs storyline generation.
The storyline includes a red agent that is tasked to complete a set of objectives in the network.
To complete the objectives, the agent needs to gain access to restricted areas of the network by exploiting and getting capabilities over the components and connectors in it.
The storyline generation process is shown in Fig.~\ref{Fig:StorylineGen}, it comprises two parts: (i) high-level storyline generation and (ii) low-level
quest generation. The first part generates a high-level storyline that spans the topology of the generated network including objectives to be completed on its subsystems.
It receives as input the output of the topology generator, i.e., a topology graph.
The second part assigns the required tasks, for each objective to be completed or for any restricted area to become accessible in each subsystem.
Therefore, it receives as input
the high-level storyline and
the subsystem for which to generate the tasks.
%The result of the storyline generation process is 
%a cybersecurity exercise scenario that can be either simulated or virtualized by being passed to either
%our simulator or by following the realization process
%described in the next section.

\begin{figure}[]
  \centering
  \includegraphics[scale=0.7]{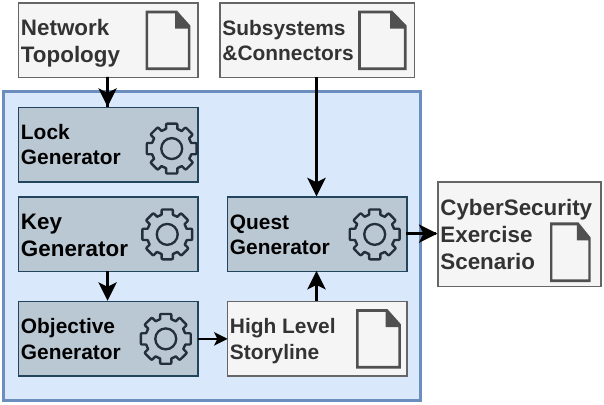}
  \caption[justification=centering]{The storyline generation process}
  \label{Fig:StorylineGen}
\end{figure}

\subsection{High Level Storyline}
The high-level storyline generation comprises two parts: (i) lock and key generation and (ii) objective generation.

\subsubsection{Lock and Key Generation}
Introducing locks restricts the access to certain parts of the network,
i.e., those behind locks, unless the attacker has performed a certain
set of actions to overcome the restriction, i.e., ``has acquired the key to the lock''. A lock can be represented by different kinds of access tokens, for example, SSH keys, user credentials or access tokens.
The lock generation algorithm is shown in Algorithm~\ref{Alg:LockGen}. It receives as input
a topology graph and the number of locks to add to it. The algorithm begins by making a copy
$C_{c}$ of the set of all connectors in the topology (line 1).  A check to ensure the number of locks
to be placed does not exceed the number of connectors follows (line 2).
This check guarantees that there exist enough connectors to not have an edge between two locks.
If the check fails the number of locks to be set is reduced to guarantee this condition (line 3).
Algorithm~\ref{Alg:LockGen} then iteratively adds $N_{\lambda}$ locks
by first picking a connector $c$ to add a lock to at random from the set of candidates and removing
the connector from the set of candidates (lines 6 and 7), next the edges of $c$ are collected in $L_{c}$
and a subset of them $l_{c}$ is selected at random to connect to the lock (lines 8 and 9), finally a
new lock is added (line 10), connected to $c$ (line 11) and the nodes in $l_{c}$ are redirected to
the new lock after being disconnected from $c$ (lines 12-16). Once the algorithm terminates,
the resulting graph is a topology with locks map, $T_{\Lambda}$ defined  %defined as follows:  = \langle S \cup C \cup \Lambda, L \cup %L_{\Lambda} \rangle$
by expanding a topology graph a set of lock nodes $\Lambda$ and a set of edges between locks and connectors $L_{\Lambda}$.

\begin{algorithm}[]
\DontPrintSemicolon 
\caption{Generate Locks}
\label{Alg:LockGen}
\footnotesize
  \KwIn{$G_{T} = \langle S \cup C, L \rangle $, a topology graph}
  \KwIn{$N_{\lambda}$, the number of locks to generate}
  \tcp{\scriptsize{Make a copy of the set of connectors (candidates)}}
  $C_{c}$ = make\_copy(C) \;
  \If { $N_{\lambda} > C$ } {
    $N_{\lambda} = C $ \;
    }
  \For{i = 0 ; i < $N_{\lambda}$; i++}{
    \tcp{\scriptsize{Pick a connector at random}}
    $c$ = random\_choice($C_{c}$) \;
    \tcp{\scriptsize{Remove connector from candidates}}
    $C_{c}$ = $C_{c} \setminus \{ c \}$  \;
      $L_{c}$ = edges\_of($T,c$) \;
      \tcp{\scriptsize{Pick a random subset of edges}}
        $l_{c}$ = random\_subset($L_{c}$) \;
      \tcp{\scriptsize{Add a lock between connector and selected edges}}
      add\_node($T$,$\lambda_{i}$) \;
      \tcp{\scriptsize{Connect lock to connector}}
    add\_edge($T$,$\lambda_{i}$,$c$) \;
    \For{ $l \in l_{c}$ }{
      \tcp{\scriptsize{Remove old edge to connector}}
      $L = L \setminus l$  \;
      \tcp{\scriptsize{Add new edge to lock}}
      $c_{t}$ = edge\_to($l$)\;
      $L = L \cup (\lambda_{i},c_{t}) $ \;
    }
  }
\end{algorithm}

The key generation algorithm is shown in Algorithm~\ref{Alg:KeyGen}, it receives as input a topology
plus locks graph and returns a topology with locks and keys graph. 
Conceptually, Algorithm~\ref{Alg:KeyGen} operates in three phases.
In the first phase, the lock nodes and edges are removed (lines 1 and 2) which results in $T_{\Lambda}$
being disconnected. In the second phase, all the connected component fragments of $T_{\Lambda}$ are
collected into a set (line 3), one of them $c$ is picked at random to place the scenario's starting
point in (line 4) and a random subsystem $s$ in $c$ is chosen to be the starting point (lines 5-8).
$c$ is then removed from the set of connected components and added to a set of merged components
$merged$. In the third phase Algorithm~\ref{Alg:KeyGen}, iteratively places $N_{k}$ keys one at a time
as follows: (i) it finds all locks $\Lambda_{C}$ that have edge(s) to a node in $merged$, picks one of them $\lambda$ at random and adds it and any edges with nodes belonging to $merged$ to $merged$
(lines 12-17), (ii) it finds all subsystems in merged, picks one at random and adds the key to
$\lambda$, $\mathsf{k}_{\lambda}$ to it (lines 18-21) and (iii) identifies all other connected components
that $\lambda$ connects to and adds them to $merged$ (lines 22 and 23). Once has completed, the
resulting graph is a topology graph plus keys and locks, $T_{\Lambda\mathsf{K}}$.
% = \langle S \cup C \cup \Lambda \cup \mathsf{K} \cup \{Start\}, L \cup %L_{\Lambda} \cup L_{\mathsf{K}} \cup \{l_{Start}\}  \rangle $.
$T_{\Lambda\mathsf{K}}$ is a topology with locks graph where the set of nodes is extended with a set of
keys $\mathsf{K}$ and a start node $Start$, and the set of edges is extended with a set of edges from
keys to the subsystems they are placed in, $L_{\mathsf{K}}$, and an edge from $Start$ to the subsystem the exercise begins from, $l_{Start}$.

\begin{algorithm}[]
\DontPrintSemicolon 
\caption{Generate Keys}
\label{Alg:KeyGen}
\footnotesize
  \KwIn{$T_{\Lambda} = \langle S \cup C \cup \Lambda, L \cup L_{\Lambda} \rangle $, a topology plus locks graph}
  \KwIn{$N_{k}$, the number of keys to generate}
  \tcp{\scriptsize{Phase1}}
  remove\_edges($T\Lambda$,$L_{\Lambda}$)\;
  remove\_nodes($T\Lambda$,$\Lambda$)\;
  CC = find\_connected\_components($T$)\;
  \tcp{\scriptsize{Phase2}}
  c = random\_choice(CC)\;
  $S_{c}$ = $S$ $\cap$ c \;
  
  s = random\_choice($S_{c}$)\;
  add\_node($Start$) \;
  add\_edge($Start$,s)\;
  CC = CC $\setminus$ \{c\}\;
  merged = c\;

  \tcp{\scriptsize{Phase3}} 
  \For{i = 0 ; i < $N_{k}$; i++}{
    
    C$_{\Lambda}$ = $\{ l \in L_{\Lambda} \mid edge\_to(l) \in merged\}$ \;
    $\lambda$ = random\_choice(C$_{\Lambda}$)\;    
    merged = merged $\cup$ $\{\lambda\}$ \;

    %\tcp{\scriptsize{Find all edges from $\lambda$ to a node in merged}} 
    \For{$l \in edges\_of(\lambda) \cap merged $}{
      add\_edge($T$,$l$)\;
    }
    %\tcp{\scriptsize{Pick a subsystem in merged}}
    $S_{k}$ =  S $\cap$ merged \;
    $s_{k}$ = random\_choice($S_{k}$)\;
    add\_node($\mathsf{k}_{\lambda}$)\;
    add\_edge($T$,$\mathsf{k}_{\lambda}$,$s_{k}$)\;
    
    %\tcp{\scriptsize{Add all connected components with an edge to the lock to merged }}
    \For{$\{ mc \in CC \mid  \exists l \in edges\_of(\lambda) \wedge edge\_to(l) \in mc  \}$}{
      merged = merged $\cup$  mc \;
    }
  }
\end{algorithm}

\subsubsection{Objective Generation}
The final elements required to complete the high-level storyline are objectives.
Objectives signify the adversary's goals within the scenario and correspond to a set of actions
that when performed the adversary's goals are achieved.  The process of generating objectives is given
in Algorithm~\ref{Alg:ObjectiveGen}. It receives a topology plus locks and keys graph and a number of
objectives to be added and returns a storyline graph.
Similarly to the lock generation algorithm the number of objectives
is bounded to the number of subsystems, thus Algorithm~\ref{Alg:ObjectiveGen} checks the number passed
as input and adjusts it to be within the allowed range (lines 1-3). Next all subsystems in $S$ are
ranked by invoking a ranking function rank\_subsystems (line 4).  $C$ is a list where the occurrences
of each subsystem are proportional to the inverse of their ranking, i.e., the higher in the ranking a
subsystem is the more times it is included in $C$. Objectives are iteratively added to the graph by:
selecting a subsystem $s$ in $C$ at random (line 6), adding an objective to it (lines 7 and 8) and
removing all occurrences of $s$ in $C$ (line 9). The resulting storyline 
graph $ \mathcal{H}$
%=  \langle S \cup C \cup \Lambda \cup \mathsf{K} \cup \mathsf{O} \cup %\{Start\}, L \cup L_{\Lambda} \cup L_{\mathsf{K}} \cup L_{\mathsf{O}} %\cup \{l_{Start}\} \rangle$. The storyline graph is an
is an extension of the topology plus locks and keys graph with objectives. In other words, the set of nodes has been expanded with a set of objectives $\mathsf{O}$ and the set of edges has been expanded with a set of edges $L_{\mathsf{O}}$ that contains an edge from each objective to the subsystem where the objective
can be completed in.

\begin{example}
  To demonstrate how the three storyline generation algorithms work we show how to apply them to the network graph in Figure~\ref{fig:network.topologies}(a) with arguments $N_{\lambda} = 1$, $N_{k} = 1$ and $N_{o} = 1$ respectively. Algorithm~\ref{Alg:LockGen} first needs to select a connector and a subset of its edges at random.
  Let us assume that the bottom connector, and the edges from it to the other two connectors were selected. Next, a new lock node is added, the selected 
  subset of edges is removed and a new set of edges from the connectors to the lock node is added. The resulting topology with locks map is shown in
  Figure~\ref{fig:storylinea}. To place the key, Algorithm~\ref{Alg:KeyGen} first removes the lock node from $G_{\Lambda}$, resulting in the disconnected graph shown in Figure~\ref{fig:storylineb}, then selects one of the two disconnected  components to add the entrypoint and the key to and subsequently
  adds back the lock connecting the two disconnected components back together.
  Let us assume that the top connected component was selected, the left subnet was selected to place the entrypoint and the right subnet was selected to place the key.
  The resulting topology with locks and keys graph $T_{\Lambda K}$ is shown in
  Figure~\ref{fig:storylinec}. Note that no matter which of the two connected graphs would be selected, since the key would be placed on one of its subsystems,
  the key in the resulting graph is guaranteed to be accessible from the entrypoint. Had the bottom connected component been selected,
  both the key and the entrypoint would be placed in the bottom subsystem.
  Finally Algorithm~\ref{Alg:ObjectiveGen} places 
  an objective by first ranking the subsystems and then by performing a biased sampling based on the score. In this case the ranking function scores subsystems that do not already contain a key and are not the entrypoint higher than subsystems that do contain a key. Thus the objective is placed in the bottom
  subsystem.
  The resulting high-level storyline graph $\mathcal{H}$ is shown in Figure~\ref{fig:storylined}.
 \end{example}

\begin{figure}
  \subfloat[$G_{\Lambda}$]
  {
  \label{fig:storylinea}
  \includegraphics[scale=0.4]{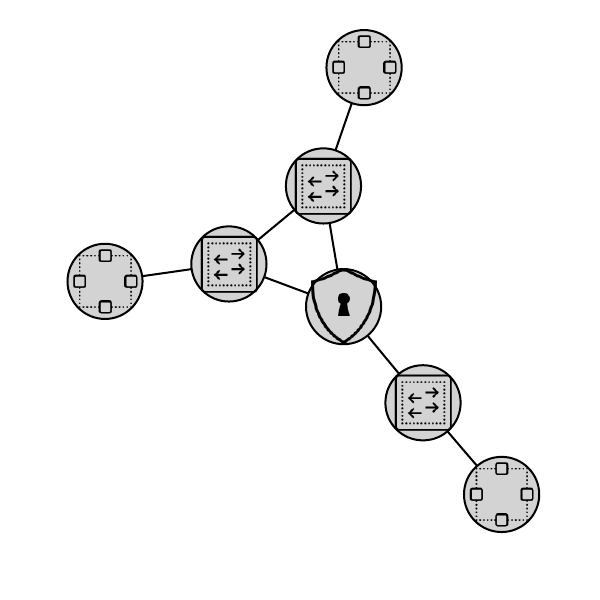}
  }
  \subfloat[Disconnected $G_{\Lambda}$]
  {
  \label{fig:storylineb}
  \includegraphics[scale=0.4]{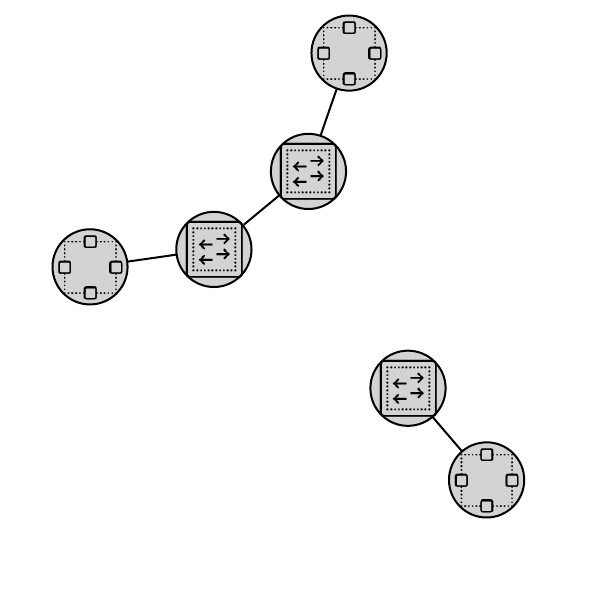}
  }\\
  \subfloat[$G_{\Lambda K}$]{\includegraphics[scale=0.37]{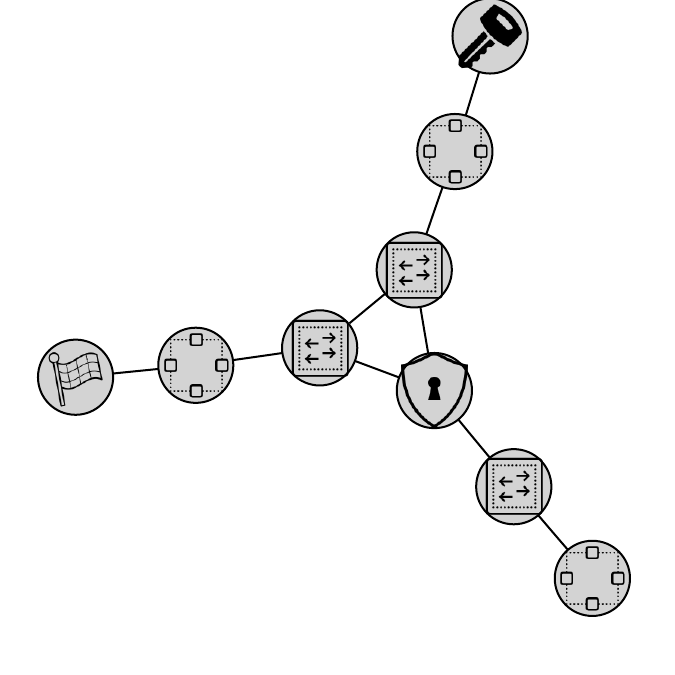}\label{fig:storylinec}}
  \subfloat[$\mathcal{H}$]{\includegraphics[scale=0.37]{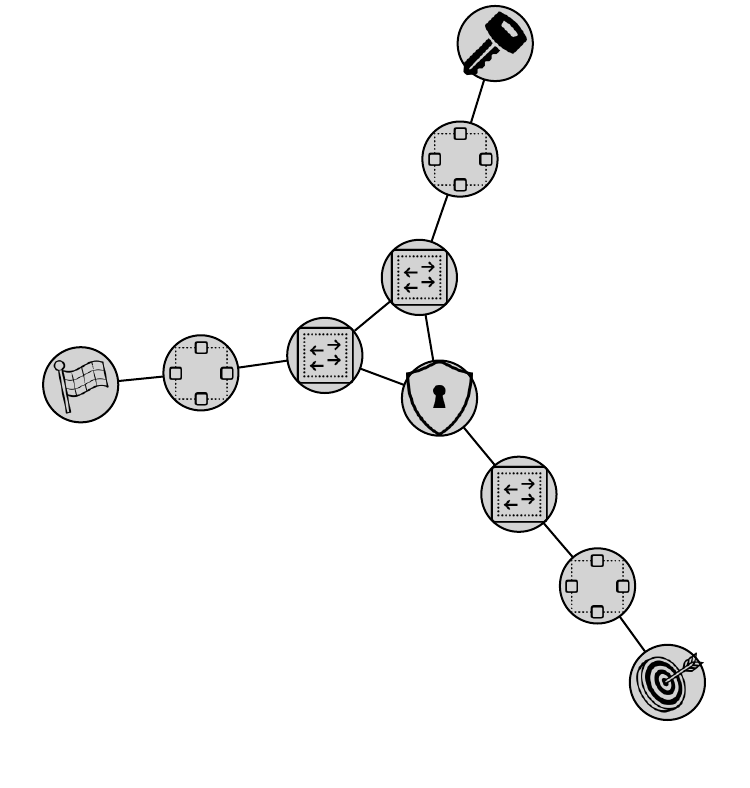}\label{fig:storylined}}
  \caption{Example1: high-level storyline generation steps}
\end{figure}

\begin{algorithm}[]
\DontPrintSemicolon 
\caption{Generate Objectives}
\label{Alg:ObjectiveGen}
\footnotesize
  \KwIn{$T_{\Lambda\mathsf{K}} = \langle S \cup C \cup \Lambda \cup \mathsf{K} \cup \{Start\}, L \cup L_{\Lambda} \cup L_{\mathsf{K}} \cup \{l_{Start}\}  \rangle $}
  \KwIn{$N_{o}$, the number of objectives to generate}

  \If{$N_{o} > |S|$}{
    $N_{o}$ = $S$\;
  }
  C = rank\_subsystems($T\Lambda\mathsf{K}$)\;
  \For{i = 0 ; i < $N_{o}$; i++}{
    s = random\_choice(C)\;
    add\_node($T\Lambda\mathsf{K}$, $\mathsf{o}$)\;
    add\_edge($T\Lambda\mathsf{K}$, $\mathsf{o}$,s)\;
    remove\_all(C,s)\;
  }
\end{algorithm}

\subsection{Low Level Storyline}
The low-level storyline is generated by making use of the
high-level storyline and the detailed subsystems and connectors representation.
We generate a low-level storyline graph per subsystem
in the network. In a low-level storyline graph of a subsystem $s$ $ \mathcal{L}_{s}$, 
 the nodes correspond to the components $C$, their interfaces, $I$ and corresponding vulnerabilities, $V$, in addition to the keys and objectives assigned to $s$, $K$ and $O$ respectively and finally an entry point marker node $E$. The set of edges is defined by edges that: (i) link communicating components ($L_{C}$), (ii) each interface with the component that contains it ($L_{I}$), (iii) each vulnerability with the interface that contains it ($L_{V}$), (iv) each key to the component in which it can be collected, (v) each objective to the component in which it can be completed and (vi) an edge that connects the entrypoint marker to the component that is initially accessible to the adversary in $s$, provided that the adversary holds all keys required to reach $s$ in the
high-level storyline graph.

\paragraph{Capabilities}
Capabilities are used in two ways: (i) to specify the capabilities required to interact with each component's interface or exploit a vulnerability and
(ii) to specify which capabilities are gained by a successful exploitation of a vulnerability or some
other action. To compute the required capabilities of an interface, we look up the Interface
parts of the architectural style definition and add a capability that signifies that the containing component is reachable. For example, given the specification in
Figure~\ref{fig:archstylespec}, and the subsystem graph shown in
Figure~\ref{Fig:SubsystemGraphExample}, the set of capabilities required to reach or exploit
the sudo interface of $Component0$ includes reaching Component0 and gaining initial access on it.

\paragraph{Quests}
We unify the storyline elements of objectives and keys through the notion of quests. 
To succeed in an objective or to collect a key,
that the adversary must complete a quest that has the corresponding objective or key as a reward. 
A quest is a series of tasks that the attacker can perform to gain a set of rewards.
Each task has a  set of prerequisites expressed as capabilities that the adversary must hold to be
able to perform and confers a set of capabilities as a reward for completion. To complete a quest,
not all of the tasks need to be completed, i.e., tasks are arranged in a partial order.
A task corresponds to a set of actions that an agent
can take to increase their capabilities in the execution environment or to reduce the capabilities of another agent that shares the same environment.
An action is an atomic observable event generated by an agent's interaction with some element of the execution environment.

\subsubsection{Assigning Keys and Objectives to Subsystems}
Before assigning keys and objectives to subsystems, 
the maximal set of capabilities over the execution environment needs to be computed. This is performed by a fairly straightforward fixpoint algorithm, the details of which
can be found in Appendix~\ref{sec:maxcap}.
Once the maximal set of capabilities over the environment have been computed, we can select a component
node $c$ and an obtainable  capability $k_{g}$ on that component at random and create a quest that
culminates in obtaining a key or completing an objective. To create a quest, we first create a fresh
capability $k_{f}$ to be appended to the set of rewards and then introduce a task $t$ that requires
$k_{g}$ to be completed and provides $k_{f}$ as a reward upon completion. To complete quest generation
we compute the partial order of all tasks that can lead to $k_{g}$ and also add $t$ to the partial
order. Once the quest generation is completed, we add a new node, either a key node or an objective
node depending on the quest type to the subsystem graph and an additional edge from the newly added
node to $c$.

\begin{figure}
  \centering
  \footnotesize
  \subfloat[$\leq$ $q_{\kappa}$]{
    \begin{tikzpicture}
      \node (max) at (5,0) {$\{t_{0}, t_{1}, t_{3}, t_{4}, t_{5}\}$};
      \node (t4) at (3,0) {$\{t_{0}, t_{1}, t_{3}, t_{4}\}$};
      \node (t3) at (1.5,0) {$\{t_{0}, t_{1}, t_{3}\}$};
      \node (t1) at (0,0) {$\{t_{0}, t_{1}\}$};
  \node (t0) at (-1,0) {$\{t_{0}\}$};
  \node (min) at (-2,0) {$\varnothing$};
  \draw (min) -- (t0) -- (t1) --(t3) -- (t4) -- (max);

\end{tikzpicture}
    }
    \hspace{4em}
\subfloat[$\leq$ $q_{o}$]{
\begin{tikzpicture}
  \node (max) at (4,0) {$\{t_{0}, t_{1}, t_{2}\}$};
  \node (t1) at (2,0) {$\{t_{0}, t_{1}\}$};
  \node (t0) at (0,0) {$\{t_{0}\}$};
  \node (min) at (-2,0) {$\varnothing$};
  \draw (min) -- (t0) -- (t1) -- (max);
\end{tikzpicture}
  }
  \caption{Example2 task partial orders}
  \label{fig:exampleLposets}
\end{figure}
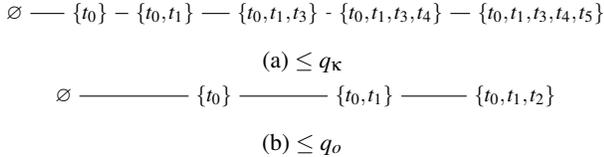

\begin{example}
  As an example, we describe how  to add a key $\kappa$ an objective $o$ and generate the corresponding quests $q_{\kappa}$ and $q_{o}$ for the subsystem shown in
  Figure~\ref{Fig:SubsystemGraphExample} with the entrypoint being Component2. After running algorithm~\ref{Alg:MaximalCaps},
  the set of obtainable capabilities includes:
  reaching Component2, gaining initial access on Component2, reaching Component1, leaking information for Component1 and reaching Component0.
    Let us assume that leaking information from Component1  and gaining initial access on Component2 are picked for the key capability Key0 and Objective0 respectively. Let us further assume the following tasks, a subset of all tasks available:
    \small{
    \begin{gather*} 
      t_{0} = \langle \{ \varnothing \}, A_{0}, \{ {Reachable\_Component2} \} \rangle \\
      t_{1} = \langle \{ {Reachable\_Component2} \}, A_{1}, \{ {InitialAccess\_Component2} \} \rangle \\
      t_{2} = \langle \{ {InitialAccess\_Component2} \}, A_{2}, \{ {Objective0} \} \rangle \\
      t_{3} = \langle \{ {InitialAccess\_Component2} \}, A_{3}, \{{Reachable\_Component1} \} \rangle \\
      t_{4} = \langle  \{{Reachable\_Component1}\}, A_{4}, \{ {InformationLeak\_Component1} \} \rangle \\
      t_{5} = \langle \{ {InformationLeak\_Component1} \}, A_{5}, \{ {Key0} \} \rangle 
\end{gather*}
}
    Figure~\ref{fig:exampleLposets} shows the corresponding quest partial orders and 
    Figure~\ref{fig:exampleLL} shows the resulting low-level storyline graph.
\end{example}

\begin{figure}
  \centering
  \includegraphics[scale=0.4]{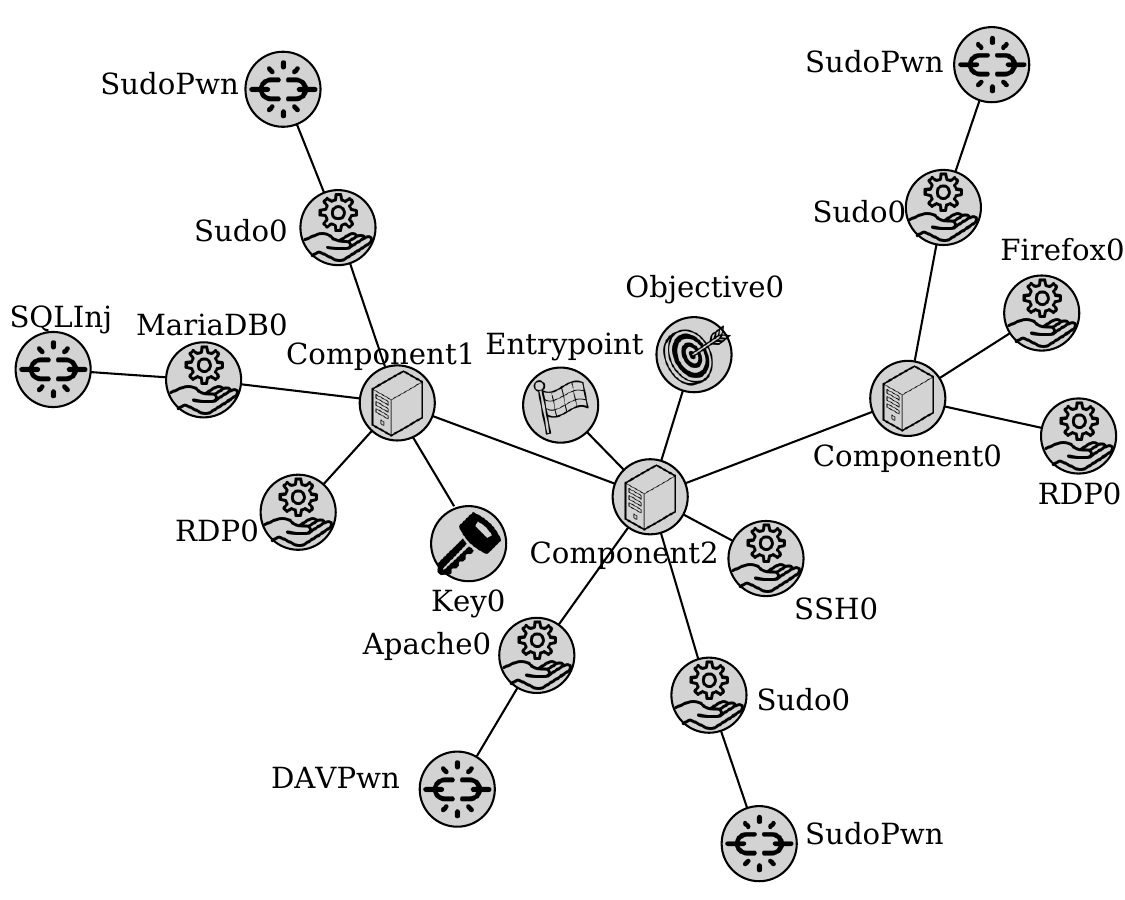}
    \caption{Example storyline graph generation} %\todo[inline]{Put the correct graph with objectives here}}
  \label{fig:exampleLL}
\end{figure}

\section{Scenario Realization}
\label{sec:scenariorealization}
To demonstrate and evaluate our scenario generation approach, we have implemented a simulated 
and a virtualized environment that can take as input a scenario and run a simulated cyberexercise in the case of the simulator, or provide the environment and partially the monitoring required to run a virtualized cybersecurity exercise. 

\subsection{Virtualized Environment}
The virtualized environment provides a more realistic platform to run cyberexercises on, with the downside of reduced traceability of agent actions which require monitoring and and event tracing mechanisms. 
%We do not provide those mechanisms as we consider them out of scope for this work.

\paragraph{Instantiating Subsystems and Connectors}
To instantiate a subsystem, we create a routed libvirt network. 
Each component in the subsystem corresponds to
a libvirt domain (virtual machine). For every interface of a component,
we provide a configuration script that installs and deploys
the corresponding interface.
Similarly, we provide configuration scripts that perform the required actions to deploy and perform any setup required for keys, objectives and vulnerabilities.
Each connector corresponds to a firewall virtual machine.
Connectors share a common routed network. Connectors that act as a bridge to a subsystem also have a (virtual) network interface attached the subsystem's network.

\paragraph{Keys, Locks and Objectives}
Keys materialize as different kinds of access tokens. For instance, they could be SSH keys or credentials for a router that allow the red team agent to change the routing rules to add access to a subsystem behind the lock or they could be SSH keys, credentials or access tokens for one or more components in a subsystem behind a lock.
The exact nature of the keys depends on the interfaces available in the subsystems behind a lock.
For each candidate interface that is able to host a key, a script must be provided to generate the key. In the default case, i.e., when no information is provided or when no interface can provide a key mechanism we fallback to generating SSH keys or credentials for the router that provides access to the subsystem behind the lock. Depending on their nature, keys can be found as files, database or registry entries etc.
We do not materialize locks explicitly and instead depend on firewall rules and interfaces that act as authorization mechanisms to provide that functionality. Objectives materialize as flags which are found
as files, database or registry entries on an exploited component's VM or can be hidden in executables or services.
They are retrieved through exploitation or by gaining the required privileges.
%For every objective a script  must also be provided.

\subsection{Simulation Environment}
Our simulator was inspired by simulators developed to train autonomous agents in penetration testing and defense tasks like Nasim~\cite{schwartz2019nasim} and CyBORG~\cite{fernandes5208526autonomous}.
Similar to the approach taken in them, the simulator can provide both a
high-level simulation environment, where only an abstracted network and high-level goals are simulated and a more detailed simulation where a detailed 
view of the network and low-level goals are simulated. Similarly to CyBORG~\cite{cage_cyborg_2022},
we simulate both a red-team agent and a blue-team agent. 
The simulation is a two-player turn-based game where iteratively on each turn, and agent performs an action on the environment, observes its effects and passes the turn to the next agent.
The red agent's goal is to complete all objectives while the blue agent's 
goal is to prevent all objectives from being completed. Each action taken
by each agent comes with an associated cost and both agents start with an initial budget from which the cost of each action is deducted.
The game ends when the red agent has succeeded in all objectives or when its budget reaches zero.
The need to implement our own simulator arose from two main limitations of existing simulators:
(i) the lack of dynamicity/flexibility in modifying firewall and routing rules during simulation
and the lack of detailed support for capabilities, key, objectives and mitigations.

\paragraph{High-level Simulation}
The high-level simulation entails simulating 
the topology and handling quests as the unit of simulation action.
Quests can succeed or fail based on a probability $p$.
The following actions are available to the red-agent:
(i) explore the network, which returns the set of susbystems that the red agent can act upon.
A red agent can freely move in the topology, adhering to firewall rules but 
cannot pass through locks unless they hold the capability associated with the
lock's key.
(ii) investigate a subsystem $s$, which will return whether and which objectives and keys are available in $s$.
(iii) KeyQuest, i.e., attempt to gain a key located in a subsystem.
(iv) ObjectiveQuest, i.e., attempt to complete a quest located in a subsystem.
The following actions are available to the blue-agent:
(i) monitor, which notifies the blue agent of any successfully completed quests on a subsystem $s$, for each completed quest this action might report incorrectly that the quest was not completed with a probability $p$.
(ii) reset a lock, which resets that state of a lock to locked,
forcing the red-team agent to reacquire the key to complete objectives behind the lock.
(iii) reset an objective, which resets the state of an objective, forcing the red agent to complete the objective again if this objective is a pre-requisite for another objective that is yet to be completed.    
Additional information is required for the high-level simulation to run, namely: 
(i) the probability of each quest succeeding or failing, (ii) the probability of false detection of quest completion for each quest, (iii) the cost of each action and (iv) the starting budgets for the agents.

\paragraph{Detailed Simulation}
The detailed simulation entails simulating both the topology and the subsystems and handing tasks as the unit
of action. Similarly to quests in the high-level simulation tasks can fail with a probability $p$.
The red-team agent propagates through the network by exploiting components and pivoting from one component 
to the next. The red agent can reach a component from another if they have at least a capability equivalent to initial access.
The actions available to the red agent include: (i) a set of scans to gain information about the topology (subsystem, connectors), (ii) a set of scans to gain information about a subsystem(components, interfaces, keys),
(iii) exploiting an interface,(iv) searching for a key,(v) achieving an objective.
The actions available to the blue team agent include: (i) checking if a component has been compromised, 
(ii) resetting a component, forcing the attacker to re-exploit its vulnerabilities if the component is required
to complete a quest that leads to gaining an objective, (iii) resetting a lock, forcing the attacker to reacquire the key, and (iv) resetting an objective, forcing the attacker to complete it again in the case that the objective is required to complete another objective not yet completed.
Similarly to the high-level simulation, the probabilities and costs of tasks need to be provided alongside
the budget for each agent as additional information.

\section{Evaluation}
\label{sec:evaluation}
We evaluate our generator in terms of of its performance and with
respect to the extend to which it satisfies the properties we identified in the introduction, namely
extensibility, modularity, solvability, 
scalability,  difficulty and variety.
We omit the evaluation of compatibility as the generator does not directly generate infrastructure.
Nevertheless, we believe that our virtualized environment does meet the compatibility definition as it is built on 
libvirt which is open source and widely available.
In the experiments that follow, we use the network generated by the specification in Figure~\ref{fig:topologyspec} but vary the numbers of subsystems.
The specification defines three kinds of architectural styles given in Appendix~\ref{Alloy}. All three styles have 3 component roles 6-7 interfaces and 6 vulnerabilities each.

%Metasploitable2 is a intentionally vulnerable VM running Ubuntu 8.04.
%It exposes multiple vulnerable services including various non-password protected services(smtp, rlogin, rsh, rexec, nfs), various backdoors(vsftpd, unrealIRCd),
%services with weak passwords(ssh, telnet, postgres, vnc), and services with well known vulnerabilities(httpd, javarmi, samba). Moreover, since the kernel version is outdated,
%there are multiple privilege-escalation exploits available once local user access has been obtained.
%Similarly, Metaspoitable3 has been designed and is available for the same purposes albeit featuring a higher degree of difficulty.
%It comes in two flavors, as a Windows windows2008 or as an Ubuntu 14.04 server. we have opted to use the Windows version in our evaluation.
%A multitude of vulnerable services, or services with weak passwords are also exposed by Metasploitable3, including for instance, struts, glassfish, mysql, iis-ftp, iis-http, etc.

\subsection{Performance evaluation}
The main performance characteristic that we are interested in is generation time.
We measure the total execution time
in addition to the time taken for each part of our approach.
Table~\ref{tab:performance} shows the time required to 
complete each step of the generation of 25 to 125 networks in addition to the total time for the whole process. Topology
and subsystem generation require a similar amount of time
and are independent of the number of networks generated.
Their contribution to total generation time is negligible.
Generating the high-level storyline($\mathcal{H}$) dominates generation time
taking up to 84-97\%. The low-level storyline generation($\mathcal{L}$) seems
to increase linearly taking up a small part of the total time
(2.5-10\%).
\begin{table}[htb]
  \caption{{Generation Time (s)}}
  \label{tab:performance}
  \centering
  \scalebox{0.7}{
    \let\Tabular\tabular
    \def\tabular{\Tabular}
    \begin{tabular}{|c|c|c|c|c|c|}
      \hline
      \textbf{\#Networks} & \textbf{Topology} & \textbf{Subsystem} &  \textbf{$\mathcal{H}$} & \textbf{$\mathcal{L}$} & \textbf{Total} \\
      \hline
      25  & 1.254 & 1.299 & 38.694 & 4.832  & 46.080\\
      \hline
      50  & 1.2479 & 1.317  & 156.425 & 9.808 & 168.799 \\ 
      \hline
      75 & 1.264 & 1.305 & 349.753 & 14.954 & 367.277\\ 
      \hline
      100 &  1.253 & 1.306 & 617.115 & 19.728 & 639.404\\ 
      \hline
      125 & 1.306 & 1.338 & 972.499 & 25.120 & 1000.260\\ 
      \hline
    \end{tabular}
  }
\end{table}

\subsection{Property Evaluation}

\paragraph{Extensibility}
To roughly estimate the extensibility of our approach, i.e., how easy it is to add and integrate new elements, we measure the number of lines of specification required to add new elements provided that their sub-elements have already been  defined. 
In particular, we consider the lines that need to be added or modified to add (i) a new vulnerable interface, (ii) a new component role and its corresponding features, and (iii)  a new topology.
Adding a new vulnerable interface requires one line for the
interface specification and one line per vulnerability specification. Adding a new component role and its features requires one line for the role specification, one line per
feature specification and one line for each of the role and feature constraints. Adding a new topology requires one line for the network topology, one line per subsystem definition and one line per subsystem constraint.
In our view, our generator is extensible as it takes a few lines of specification to add any new element. Further external evaluation would be needed to properly assess the true extensibility.

\paragraph{Modularity}
Our approach is modular in the sense that once elements have been defined they can be freely reused to generate different
networks. In particular, the following elements can be reused and composed
independently once defined: (i) vulnerable interfaces, (ii) component roles and their features (iii) subsystems. This meets the stated goal of being able to reuse and compose elements without customization.

\paragraph{Solvability}
For the simulated environments the solvability of our approach
is guaranteed by construction due to the following properties
of algorithm~\ref{Alg:KeyGen} and algorithm~\ref{Alg:MaximalCaps}. For the virtualized   environment,
the solvability of our approach is guaranteed by the same algorithms modulo the existence of a working exploit for each vulnerability and non-destructiveness, i.e., exploits do not 
have adverse side-effects that can cause network connections to 
be disrupted or hosts to go down etc.

Algorithm~\ref{Alg:KeyGen} guarantees that each key is placed in a subsystem that can be reached before the corresponding lock. Therefore, all locks can be unlocked in both the high-level and the low-level storyline. Algorithm~\ref{Alg:KeyGen} begins by disconnecting 
the topology with locks graph by removing all of the nodes
and edges that correspond to locks resulting in a number of connected components that are not accessible from one another. In the first iteration, the algorithm picks one of the connected components at random and assigns one of its subsystems as the entrypoint. While the choice is random, any chosen subsystem is a valid option that maintains the required invariant of placing the keys in accessible areas of the graph before the locks. This is guaranteed by placing the key in a subsystem
of the currently chosen connected component, then placing 
its corresponding lock and expanding the connected component
by merging all of the connected components that the added lock 
"bridges" to form a larger connected component.
The process of adding keys before adding the corresponding
locks is repeated until all locks have been placed and 
all connected components have been merged.
%Note that in each step of the process the same invariant holds:
%the key is first placed in a subsystem of the accessible area and then we add the lock and merge the "bridged" connected
%component, therefore for every lock, its key can always be found in a path starting from the entrypoint that is not blocked by the lock.

Objectives and keys are also guaranteed to be placed where they can be achieved in each subsystem storyline graph. Quests are placed after running algorithm~\ref{Alg:MaximalCaps} to calculate the set of maximal capabilities in the subsystem. Since we assign their prerequisite capabilities from the set returned by algorithm~\ref{Alg:MaximalCaps}, we can be certain that the prerequisites are achievable and thus the quests can always be completed.

\paragraph{Scalability}
Table~\ref{tab:scalability} shows the number of different topologies (\#T) and networks (\#N) that can be generated by 
our approach as the number of subsystems increases. 
Each topology and network is composed of different numbers 
of the subsystem types (\#S) shown in Figure~\ref{fig:topologyspec}.
Subsystems generated from the ClientServer and Oauth styles have 3 components (\#C) , while those generated by the Microservices style have either 2 or 3 components each. The table shows
the even with fairly small numbers of subsystems and components
our approach can generate a large number of topologies and networks. 

\begin{table}[htb]
  % \begin{wraptable}[7]{R}{8cm}
  % \vspace{-\intextsep}
  \caption{{Size statistics for different numbers of subsystems}}
  \label{tab:scalability}
  \centering
  \scalebox{0.7}{
    \let\Tabular\tabular
    \def\tabular{\Tabular}
    \begin{tabular}{|c|c|c|c|}
      \hline
      \textbf{\#S} & \textbf{\#C} &  \textbf{\#T} & \textbf{\#N} \\
      \hline
      5 & 13 & 27 & 15552 \\
      \hline
      6 & 16 & 88 & 50668 \\ 
      \hline
      7 & 19 & 177 & 101952 \\ 
      \hline
      8 & 22 & 499 & 287424 \\ 
      \hline
      9 & 24 & 1037 & 597313 \\ 
      \hline
      10 & 27 & 3392 & 1953792 \\
      \hline
      11 & 29 & 7289 & 4198464 \\
      \hline
      12 & 32 & 19974 & 11505024 \\
      \hline
    \end{tabular}
  }
  % \end{wraptable}
\end{table}

\begin{figure}
\centering
\scalebox{0.6}{
\begin{tikzpicture}[]
\begin{axis}[
    xlabel = {\Large{\#Networks}},
    ylabel = {\Large{Completed Tasks}},
    xmin=0, xmax=12,
    ymin=0, ymax=60,
    xtick={3,5,7,9,12},
    scale only axis,
    legend pos=north west,
    ymajorgrids=true,
	enlarge x limits=0.05,
    grid style=dashed,
    every axis plot/.append style={thick},
    every tick label/.append style={font=\Large}
]
 
\addplot[
    color=Goldenrod,
    mark=*,
    ]
coordinates {
(3, 7.5)
(5, 9.75)
(7, 11.3)
(9, 13.8)
(12,15.2)
};

\addplot[
    color=RoyalPurple,
    mark=*,
    ]
coordinates {
(3, 17.5)
(5, 20.75)
(7, 24.2)
(9, 27.9)
(12,31.3)

};

\addplot[
    color=RoyalBlue,
    mark=*,
    ]
coordinates {
(5, 26.75)
(7, 35.25)
(9, 45.50)
(12,54.2)
};
\legend{\Large{$N_{\lambda} = 1  \mathbin{,} N_{o} = 1$},\Large{$N_{\lambda} = 2  \mathbin{,} N_{o} = 2$},\Large{$N_{\lambda} = 3  \mathbin{,} N_{o} = 3$}}
\end{axis}
\end{tikzpicture}
}
\caption[justification=centering]{Difficulty measurements}
\label{Fig:Difficulty}
\end{figure}
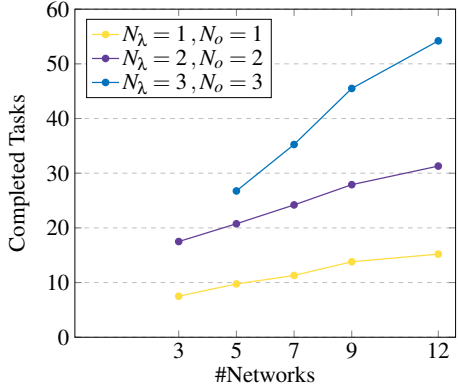

\paragraph{Difficulty}
To measure difficulty, we calculate the average number of tasks that need to be completed to achieve all objectives in networks with a different number of subsystems but also with a different number of locks and objectives present.
Figure~\ref{Fig:Difficulty} shows the average number of tasks needed to complete all objectives for networks with 3,5,7,9,12 subsystems and 1,2,3 locks and objectives respectively.
Note that for networks with only 3 subsystems, there can be at most 2 locks and 2 objectives hence there is no value for 3 locks and 3 objectives in the graph.
The results indicate, that both the number of locks and objectives,
and the size of the network influence number of tasks to be
completed. We can tune the difficulty of the generated network by selecting different networks sizes and number of locks and objectives. Therefore, 
we can indeed generate scenarios that vary in the effort
that they require to complete.

\paragraph{Variety}
To measure variety, we consider two cases: structural
variety which corresponds to the variety in the structure of the networks generated by our approach and content variety
which corresponds to the variety in the different elements included in the networks generated by our approach.

\begin{figure}
  \centering
  \includegraphics[scale=0.35]{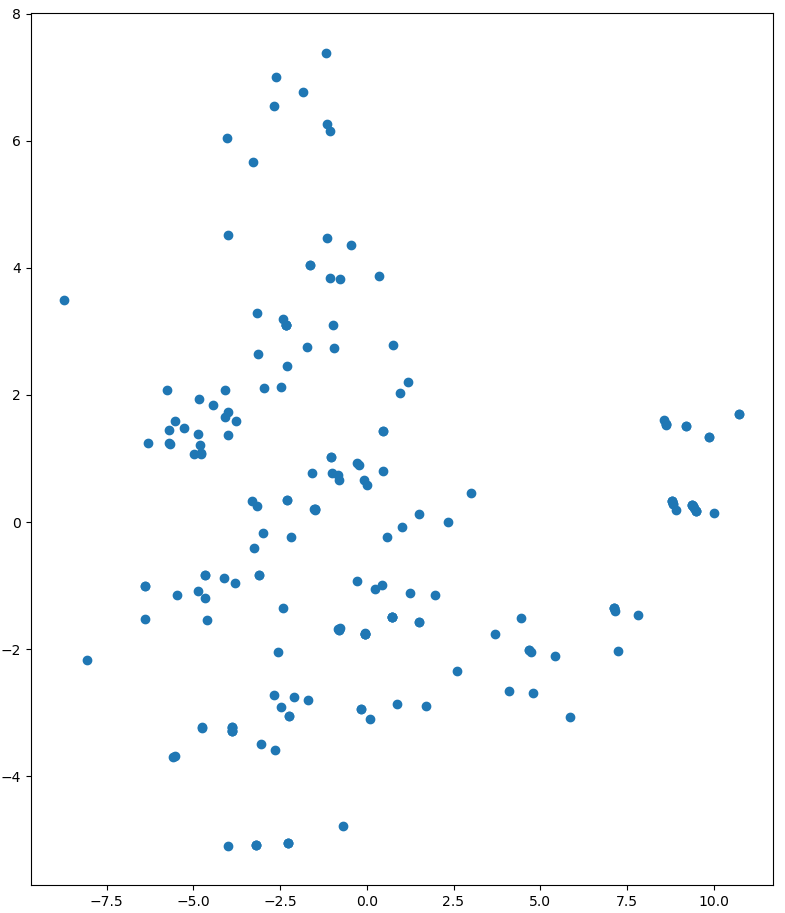}
    \caption{Scatterplot demonstrating structural variety}
  \label{fig:structurepca}
\end{figure}
To measure structural variety, we use the NetSimile similarity metric~\cite{NetSimile}. 
For each network, we first merge its topology graph with all of its subsystem graphs to get a single graph the represents the whole network. 
Next, we generate the Netsimile signature vector for each 
network graph and perform a principal component analysis
on the set of network graph signatures to reduce the dimensionality of the results to two dimensions. 
Since the first two principal components can explain over 95\% of the variance in the results we draw the scatterplot shown in Figure~\ref{fig:structurepca}. Figure~\ref{fig:structurepca} shows the results for 100 networks, we have repeated this process with networks of larger sizes getting similar results. The results indicate a high degree of structural variety across the different networks.

To measure content variety, we generate a signature vector for each network.
Each signature vector has a component that corresponds to each interface
and each vulnerability potentially present in the network according to the
specification of its subsystems. We then apply an agglomerative hierarchical clustering algorithm and produce a heatmap of the results. The results for 100 
generated networks are shown in Figure~\ref{fig:contentheatmap}. 
The results indicate a high degree of content variety.

\begin{figure}[]
\centering
  \includegraphics[scale=0.3]{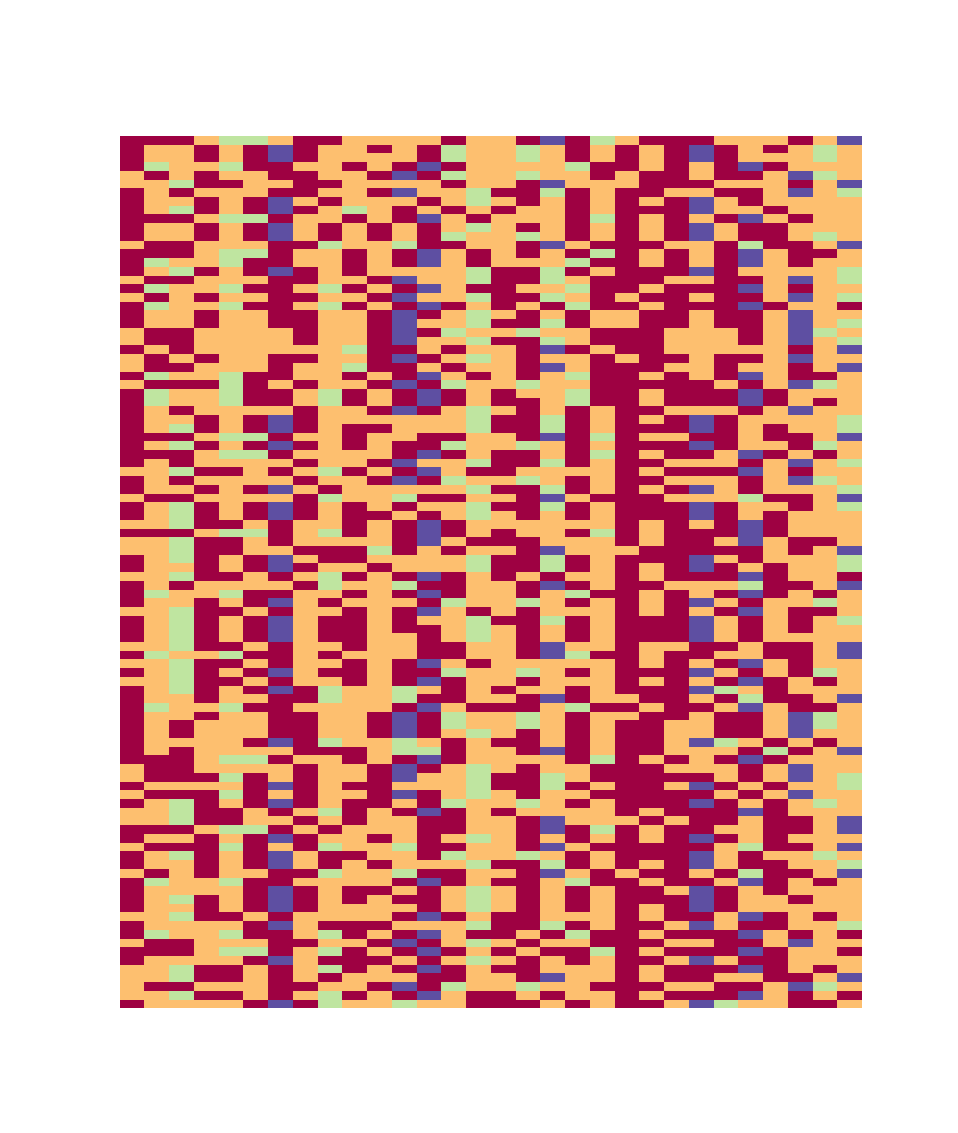}
    \caption{Interface and vulnerability heatmap}
  \label{fig:contentheatmap}
\end{figure}
\section{Related Work}
\label{sec:related.work}

Multiple scenario generation approaches targeting a specific cyberange environment can be found in the literature.
The CRACK Scenario Definition Language~\cite{RUSSO2020101837} extends the Topology and Orchestration Specification for Cloud Applications (TOSCA) to model, verify, and test cybersecurity training scenarios. It adds cybersecurity scenario-related node and relationship types, verification properties,
and a pattern language for structured queries.
The Collaborative Security Training Scenario Description Language~\cite{CST-SDL} (CST-SDL) is a model-driven approach to specify and automate the deployment of cybersecurity training scenarios. It provides collaborative training environments, supporting multi-trainee and multi-solution scenarios.
Additional model-driven approaches include: (i) the Cybersecurity Training Language~\cite{10.1007/978-3-030-42051-2_8} (CTL) and (ii) The Virtual Scenario Description Language~\cite{VSDL} (VSDL),  (iii) the scenario definition language employed by The Open Cyber Range (OCR) platform, (iv) 
EDURange~\cite{Weiss}  (v) the KYPO Cyber Range~\cite{KYPO} and
(vi) CyRIS~\cite{CyRIS}.
The Serious Game scenario description language~\cite{SERIOUS} employs gamification to design cybersecurity scenarios. Moreover, it is designed to provide an interactive platform for modeling and testing cyber-attack and defense strategies. 
The above approaches generally focus on the design of cybersecurity exercises that will be exclusively deployed on the accompanying cyberrange and thus their design elements are specific to the options available
within that cyberange. Our approach aims to be more general taking a systems approach
to design scenarios that can be deployed in both simulators and emulated environments by implementing backends similar to the ones we describe in Section~\ref{sec:scenariorealization}. Moreover, the above approaches require that scenarios are designed individually and elements are chosen specifically for each instantiated scenario, this work aims to automate the process of scenario generation based on a set of network topologies and architectural styles, this way, a large amount of scenarios can be generated from a single specification, in contrast to the one-to-one relationship between scenario specification and 
generated scenario in the above approaches.

ALPACA~\cite{ALPACA} uses a Prolog-based engine
to generate a virtual machine containing a set of vulnerabilities organized on a dependency lattice ensuring that exploits exist for all vulnerabilities. 
SecGen~\cite{205221} creates vulnerable virtual machines to be used for learning penetration testing techniques. It includes a catalog of vulnerabilities that can be randomly selected based on constraints within the scenario definition. These approaches generate a single VM and aim for penetration testing rather than cybersecurity exercises. They are similar to our approach in that they specify a repository of possible components each with a set of possible vulnerabilities that can be used to instantiate a variety of different options.

 Cyexec$^{*}$~\cite{nakata2021cyexec} employs graph randomization techniques to generate cybersecurity exercise scenarios by rearranging small pre-generated scenario fragments into a graph structure and employing milestones, i.e., goals that can be achieved through different exploitation means. Randomization amounts to picking among the different exploitation means available for each milestone present, thus generating a scenario for every combination of milestones and exploitation means. The generated scenarios have the same objectives which can be achieved in different ways. Our approach generates multiple possible networks instantiating architectural styles into multiple different subsystems and connects them in different ways that satisfy the constraints imposed by the allowed network topology. As a result, the variety of the graphs generated by our approach is higher compared to those
 generated by Cyexec$^{*}$. Moreover, our approach introduces variety in the goals and objectives by randomly generating a number of objectives that can be completed in addition to obstacles (locks) that must be overcome before the objectives are attainable.

Large language models~\cite{10695083} and machine learning~\cite{zacharis2023ai,zacharis2023aicef}
have also used to generate scenarios, however, these approaches do not generate an execution environment or a machine readable specification of 
a scenario. Instead they run similar to tabletop exercises and require additional effort to clarify the details required to generate a machine readable specification and  automate its deployment.

\section{Conclusion}
\label{sec:conclusion}

This work contributes an approach to generate a large amount of interesting cybersecurity scenarios that can be used to train agents, both human and computational in operational security.
We have implemented our approach and release our scenario generator in addition to two scenario execution environment
backends that can receive a scenario generated by our approach and can provide a simulation and a
virtualization environment respectively on which cybersecurity exercises can be run.

%As future work, we plan to work on expanding the dynamic aspects of our approach and the development of more roles, for instance, green team operators.
%A promising direction is to incorporate dynamic goal models to define both the objectives and the
%agent's success criteria. A further avenue is to support additional kinds of networks, connectors
%and components beyond those that belong to enterprise networks.

\appendix
\section{Ethical Considerations}
We acknowledge the potential that some of the outcomes of this work could be used to train  malicious (red team) agents. However, we believe that the
utility provided outweighs the risk as training red team agents is necessary to be able to analyze and study their operation
to develop mitigations and blue team agents that can counteract them.
Furthermore, with regards to the released dataset, we have intentionally opted to use older operating systems and services with vulnerabilities that have been 
openly known for a long time period and thus are unlikely to be
widely deployed.
\section{Open Science}

In the interest of open science, we will provide our dataset alongside a snapshot of the scenario generator, simulator and virtualization sources, released under an open source license upon acceptance.
\bibliographystyle{plain}
\bibliography{bib.bib}

\appendix
\section{Alloy and Architectural Style Specifications}
\label{Alloy}
To aid reproducibility, alongside the artifacts being available to download, we additionally list all the alloy 
and architectural style specifications involved in the
experiments. Moreover, we also show the translated part of Fig.~\ref{fig:archstylespec}.

 \begin{figure*}
  \centering
\begin{lstlisting}[language = Alloy]
abstract sig ArchStyle { implementedIn:some Subsystem}

sig Subsystem {
style: one ArchStyle,
router: one Connector,
}

sig Connector {
bridges: some Subsystem,
inNetwork: one Network,
}

sig Link
{
  connects : some Connector,
  
}{ #(connects) = 2 } 

abstract sig Network {
routers : some Connector
}

pred requires[as1: ArchStyle, as2:ArchStyle] {
 #as1.implementedIn > 0 iff #as2.implementedIn > 0
}

pred can_reach_directly[r1: Connector, r2: Connector]
{
	(some l: Link | r1 in l.connects and r2 in l.connects) 
}

fact world_order {
//Relations between abstract signatures
bridges = ~router
routers = ~inNetwork
style = ~implementedIn

all l1, l2 : Link | l1.connects = l2.connects iff l1 = l2

//If a style s requires style sr then there exists a path 
all s1 : Subsystem | all s2: Subsystem | all s: s1.style 
                   | all sr: s2.style | requires[s,sr] => can_reach_directly[s1.router,s2.router]

//A router bridges at least two VLANs unless there is a single Subnets
#FlatNet = 0 implies #Connector > 1 and all r: Connector | #r.bridges >= 1
#FlatNet = 1 implies  (#Connector = 1 and #Subsystem = 1)
//There is exactly one network
#Network = 1
}

//Networks
lone sig FlatNet extends Network { } 
lone sig CollapsedCore extends Network { }
\end{lstlisting}

\caption{Common part of a topology specification}

\label{TopologyCommon}
\end{figure*}

\begin{figure*}
  \centering  
\begin{lstlisting}[language = Alloy]
abstract sig Style { requiresRole : some Role, satisfyingComponents: some Component}
sig Component { Interfaces : some Interface,playsRole : one Role,inStyle: some Style}
abstract sig Role {playedBy : some Component, requiresFeature : some Feature,requiredByStyle : some Style}
abstract sig Feature {implementedIn : some Interface,requiredByRole : some Role}
abstract sig Interface {comp : some Component,implements : one Feature}
abstract sig Mapping {left : some Component,right : some Component}
abstract sig OneToOne extends Mapping {} {#left = 1 #right = 1}
abstract sig OneToSome extends Mapping {} {#left = 1 #right >= 1} 
abstract sig OneToAll extends Mapping {} {#left = 1 #right >= 1}
abstract sig SomeToOne extends Mapping {} {#left >= 1 #right = 1}
abstract sig SomeToSome extends Mapping {} {#left >= 1 #right >= 1}
abstract sig SomeToAll extends Mapping {} {#left >= 1 #right >= 1}
abstract sig AllToOne extends Mapping {} {#left >= 1 #right = 1}
abstract sig AllToSome extends Mapping {} {#left >= 1 #right >= 1}
abstract sig AllToAll extends Mapping {} {#left >= 1 #right >= 1}
pred one_to_one[m:Mapping] {
	all c0: m.left,  c1:m.left | c0.playsRole = c1.playsRole
	all c0: m.right,  c1:m.right | c0.playsRole = c1.playsRole
	#m.left = 1
	#m.right = 1}
pred one_to_some[m:Mapping] {
	all c0: m.left,  c1:m.left | c0.playsRole = c1.playsRole
	all c0: m.right,  c1:m.right | c0.playsRole = c1.playsRole
	all r: m.right.playsRole | some c0: r.playedBy | c0 in m.right
	#m.left = 1}
pred one_to_all[m:Mapping] {
	all c0: m.left,  c1:m.left | c0.playsRole = c1.playsRole
	all c0: m.right,  c1:m.right | c0.playsRole = c1.playsRole
	all r: m.right.playsRole | all c0: r.playedBy | c0 in m.right
	#m.left = 1}
pred some_to_one[m:Mapping] {
all c0: m.left,  c1:m.left | c0.playsRole = c1.playsRole
all c0: m.right,  c1:m.right | c0.playsRole = c1.playsRole
all r: m.left.playsRole | some c0: r.playedBy | c0 in m.left
#m.right = 1}
pred some_to_some[m:Mapping] {
all c0: m.left,  c1:m.left | c0.playsRole = c1.playsRole
all c0: m.right,  c1:m.right | c0.playsRole = c1.playsRole
all r: m.left.playsRole | some c0: r.playedBy | c0 in m.left
all r: m.right.playsRole | some c0: r.playedBy | c0 in m.right}
pred some_to_all[m:Mapping] {
all c0: m.left,  c1:m.left | c0.playsRole = c1.playsRole
all c0: m.right,  c1:m.right | c0.playsRole = c1.playsRole
all r: m.left.playsRole | some c0: r.playedBy | c0 in m.left
all r: m.right.playsRole | all c0: r.playedBy | c0 in m.right}
pred all_to_one[m:Mapping] {
all c0: m.left,  c1:m.left | c0.playsRole = c1.playsRole
all c0: m.right,  c1:m.right | c0.playsRole = c1.playsRole
all r: m.left.playsRole | all c0: r.playedBy | c0 in m.left
#m.right = 1}
pred all_to_some[m:Mapping] {
all c0: m.left,  c1:m.left | c0.playsRole = c1.playsRole
all c0: m.right,  c1:m.right | c0.playsRole = c1.playsRole
all r: m.left.playsRole | all c0: r.playedBy | c0 in m.left
all r: m.right.playsRole | some c0: r.playedBy | c0 in m.right}

pred all_to_all[m:Mapping] {
all c0: m.left,  c1:m.left | c0.playsRole = c1.playsRole
all c0: m.right,  c1:m.right | c0.playsRole = c1.playsRole
all r: m.left.playsRole | all c0: r.playedBy | c0 in m.left
all r: m.right.playsRole | all c0: r.playedBy | c0 in m.right}
\end{lstlisting}
\caption{Common part of an architectural style specification - part 1}
\label{ArchStyleCommonA}
\end{figure*}

\begin{figure*}
  \centering  
\begin{lstlisting}[language = Alloy]
pred unique_feature[r: Role, f: Feature] {
all c: Component | ((c.playsRole = r) => not (some i0,i1:Interface 
                 | i0 in c.Interfaces and i1 in c.Interfaces and i0.implements = f 
                 and i1.implements = f and i0  != i1))
}

pred unique_role[r:Role] {
all c: Component | (c.playsRole = r) => not (some c0: Component | c0.playsRole=r and c != c0)
}

fact world_order {
Interfaces = ~comp
playsRole = ~playedBy
implementedIn = ~implements
requiresFeature = ~requiredByRole
requiresRole = ~requiredByStyle
satisfyingComponents = ~inStyle

all r:Role | some s: Style | r in s.requiresRole
all r: Role, s: Style | r in s.requiresRole => 
    (some c: Component | c.playsRole = r and c in s.satisfyingComponents)
not some c: Component | c.inStyle = none
all i : Interface, c : Component | i in c.Interfaces =>
    (some f : Feature | i in f.implementedIn and (some r: Role | f in r.requiresFeature and r = c.playsRole))
all f : Feature , r : Role| some c : Component | (f not in r.requiresFeature) => 
    not (some i : Interface | i.implements = f and i in c.Interfaces)
all c: Component, r:Role , f:Feature | r in c.playsRole and f in r.requiresFeature =>
    (some i: Interface | i in f.implementedIn and i in c.Interfaces)
all c: Component | some m: Mapping |  c in m.left or c in m.right
\end{lstlisting}
\caption{Common part of an architectural style specification - part 2}
\label{ArchStyleCommonB}
\end{figure*}

\begin{figure*}
  \centering  
\begin{lstlisting}[language = Alloy]
ArchStyle ClientServer
Role WebServer:1
Role WebClient:3
Role Database:1
Feature WebSiteHosting:WebServer
Feature RelationalQueryHandling:Database
Feature WebsiteAccess:WebClient
Feature RunAs:WebServer,WebClient,Database
Interface ApacheHTTPD:WebSiteHosting 0
Interface Tomcat:WebSiteHosting 0
Interface PostgreSQL:RelationalQueryHandling 0
Interface MariaDB:RelationalQueryHandling 0
Interface Firefox:WebsiteAccess 0
Interface Chromium:WebsiteAccess 0
Interface Sudo:RunAs INITIAL_ACCESS
Map WebServer Database OneToOne
Map WebClient WebServer AllToOne
Constraint unique_feature WebServer WebSiteHosting
Constraint unique_feature Database RelationalQueryHandling
Constraint unique_feature WebClient WebsiteAccess
Constraint unique_role WebServer
Constraint unique_role Database
Vulnerability DAVPwn ApacheHTTPD 0 INITIAL_ACCESS
Vulnerability StrutsPwn Tomcat 0 INITIAL_ACCESS
Vulnerability PostgreSQLInj PostgeSQL 0 DATABASE_DUMP
Vulnerability MySQLInj MySQL 0 DATABASE_DUMP
Vulnerability MariaDBInj MariaDB 0 DATABASE_DUMP
Vulnerability SudoPwn Sudo INITIAL_ACCESS ROOT_ACCESS
Entrypoint WebServer
\end{lstlisting}
\caption{ClientServer Architectural Style}
\label{fig:ClientServer}
\end{figure*}

\begin{figure*}
  \centering  
\begin{lstlisting}[language = Alloy]
ArchStyle OAuth
Role AuthAPI:1
Role AuthDB:1
Role Client:2
Feature AuthUser:AuthAPI
Feature FetchUserInfo:AuthDB
Feature GetAuthToken:Client
Interface RESTAPI:AuthUser 0
Interface HttpPOST:AuthUser 0
Interface MySQL:FetchUserInfo 0
Interface MariaDB:FetchUserInfo 0
Interface PostgreSQL:FetchUserInfo 0
Interface GetHTTPToken:GetAuthToken 0
Interface GetCert:GetAuthToken 0
Map AuthAPI AuthDB OneToOne
Map Client AuthAPI AllToOne
Constraint unique_feature AuthAPI AuthUser
Constraint unique_feature AuthDB FetchUserInfo
Constraint unique_role AuthAPI
Constraint unique_role AuthDB
Vulnerability RestCommandInjection RESTAPI 0 INITIAL_ACCESS
Vulnerability MaliciousRedirect HttpPOST 0 INITIAL_ACCESS
Vulnerability MySQLInj MySQL 0 DATABASE_DUMP
Vulnerability PostgreSQLInj PostgeSQL 0 DATABASE_DUMP
Vulnerability MariaDBInj MariaDB 0 DATABASE_DUMP
Vulnerability TokenForgery GetHTTPToken 0 INITIAL_ACCESS
Vulnerability TokenSkip GetCert 0 INITIAL_ACCESS
Entrypoint AuthAPI
\end{lstlisting}
\caption{OAuth Architectural Style}
\label{fig:OAuth}
\end{figure*}

\begin{figure*}
  \centering  
\begin{lstlisting}[language = Alloy]
ArchStyle Microservices
Role APIGate:1
Role BackendService:2
Role FrontendService:2
Feature APIDelegate:APIGate
Feature BackendWorker:BackendService
Feature APIConsumer:FrontendService
Interface SpringAPIGateway:APIDelegate 0
Interface SpringMSControl:APIConsumer 0
Interface SpringMSService:BackendWorker 0
Interface GoMicroGreeter:APIDelegate 0
Interface GoMicroService:BackendWorker 0
Interface GoMicroAPI:APIConsumer 0
Map APIGate FrontendService OneToAll
Map FrontendService BackendService AllToSome
Constraint unique_feature APIGate APIDelegate
Constraint unique_role APIGate
Vulnerability SpringAPICommandInjection SpringAPIGateway 0 INITIAL_ACCESS
Vulnerability SpringCommandInjection SpringMSControl 0 INITIAL_ACCESS
Vulnerability SpringCodeInjection SpringMSService 0 INITIAL_ACCESS
Vulnerability GoMicroAPICommandInjection GoMicroGreeter 0 INITIAL_ACCESS
Vulnerability GoMicroCommandInjection GoMicroAPI 0 INITIAL_ACCESS
Vulnerability GoMicroCodeInjection GoMicroService 0 INITIAL_ACCESS
Entrypoint APIGate
\end{lstlisting}
\caption{Microservices Architectural Style}
\label{fig:Microservices}
\end{figure*}

\begin{figure}
\begin{lstlisting}[language = Alloy]
one sig ClientServer extends Style { }
one sig WebServer extends Role { }
one sig WebClient extends Role { }
one sig WebHosting extends Feature { }
one sig WebAccess extends Feature { }
one sig RunAs extends Feature { }
lone sig Apache extends Interface { }
lone sig Firefox extends Interface { }
lone sig Sudo extends Interface { }
one sig WebClientServerAllToOne extends AllToOne{}
fact world_mapping {
ClientServer.Roles = WebServer + WebClient
WebServer.Features = WebHosting  + RunAs
WebClient.Features = WebAccess + RunAs
#Apache > 0 => Apache.implements = WebHosting
#Firefox > 0 => Firefox.implements = WebsiteAccess
#Sudo > 0 => Sudo.implements = RunAs
unique_feature[WebServer,WebHosting]
unique_feature[WebClient,WebAccess]
unique_role[WebServer]
all_to_one[WebClientWebServerAllToOne] }
run {#WebServer.playedBy = 1,
     #WebClient.playedBy = 2}
\end{lstlisting}
\caption{Alloy translation of the definition in Figure~\ref{fig:archstylespec}}
\label{fig:alloyarchspec}
\end{figure}

\section{Maximal Capability Set Calculation}
\label{sec:maxcap}

\begin{algorithm}[]
\DontPrintSemicolon 
\caption{Maximal capability set calculation}
\label{Alg:MaximalCaps}
\footnotesize
\KwIn{$T$, the set of available tasks in a subsystem $S$}
\KwIn{$K_{0}$ a potentially empty initial set of capabilities}

progress = $\top$ \;
$K_{max}$ = $K_{0}$ \;
\While{progress = $\top$ $\wedge$ |T| > 0}{
    progress = $\bot$ \;
    \For{ t $\in$ T }{
    \tcp{t = $K_{*},A, K_{+}$}
    \If{$K_{*} \subseteq K_{max}$ }{
        $K_{max} = K_{max} \cup K_{+} $\;
        $T = T \setminus \{t\}$ \;
        progress = $\top$ \;
    }    
    }
}
  
\end{algorithm}

Algorithm~\ref{Alg:MaximalCaps} computes the maximal set of capabilities in a subsystem.
Starting with an initial set of capabilities, $K_{0}$,
the algorithm attempts to succeed in at least one 
task $t$ in each step of the outer loop (line 3), using the progress variable
to identify is at least one task succeeded. When a task succeeds: the capabilities gained for completing it are added to the set of maximal capabilities $K_{max}$, the task is removed from the set of possible tasks
and progress is set to $\top$ to continue looping.
The algorithm ends when a fixpoint is reached, either by all capabilities being added to $K_{max}$ or by the inner loop making no progress, which means that no new task can be completed.

\end{document}